\documentclass[aps,prd,preprintnumbers,groupedaddress,nofootinbib,showpacs]{revtex4}

\usepackage{aas_macros}
\usepackage{graphicx} 
\usepackage{amssymb}
\usepackage{amsmath}
\usepackage{subfigure}
\usepackage{color}

\begin{document}

\title{Galactic PeV neutrinos from dark matter annihilation}
\author{Jes\'us Zavala}
\email{jzavala@dark-cosmology.dk}
\thanks{Marie Curie Fellow}
\affiliation{Dark Cosmology Centre, Niels Bohr Institute, University of Copenhagen, Juliane Maries Vej 30, 2100 Copenhagen, Denmark}

\preprint{}
\date{\today}

\begin{abstract}
The IceCube Neutrino Observatory has observed highly energetic neutrinos in excess of the expected atmospheric neutrino background.
It is intriguing to consider the possibility that such events are probing fundamental physics beyond the standard model of particle physics. In this context, 
$\mathcal{O}$(PeV) dark matter particles decaying to neutrinos have been considered while dark matter annihilation
has been dismissed invoking the unitarity bound as a limiting factor for the annihilation rate. However, the latter claim was done ignoring the contribution from dark matter substructure, which in a PeV Cold Dark Matter scenario, would extend down to a free streaming mass of $\mathcal{O}$($10^{-18}$M$_\odot$). Since the unitarity bound
is less stringent at low velocities, ($\sigma_{\rm ann}$v)$\leq4\pi/m_\chi^2v$, then, it is possible that these cold and dense subhalos
would contribute dominantly to a dark-matter-induced neutrino flux and easily account for the events observed by IceCube. A dark matter model where annihilations
are enhanced by a Sommerfeld mechanism can naturally support such scenario.
Interestingly, the spatial distribution of the events shows features that
would be expected in a dark matter interpretation. Although not conclusive, 9 of the 37 events appear to be clustered around an extended region near the Galactic Center while 6 others spatially coincide, 
within the reported angular errors, with 5 of 26 Milky Way satellites. However, a simple estimate of the probability of the latter occurring by chance is $\sim35\%$. More events are needed to statistically test this
hypothesis.  PeV dark matter particles are massive enough that their abundance as standard thermal relics would overclose the Universe. This issue can be solved in alternative scenarios, 
for instance if the decay of new massive unstable particles generates significant entropy reheating the Universe to a slightly lower temperature than the freeze-out temperature, $T_{\rm RH} \lesssim T_{\rm f}\sim4\times10^4$~GeV.

\end{abstract}


\maketitle







\section{Introduction}

The IceCube collaboration has recently announced the possible detection of the first high energy neutrinos with a cosmic origin \cite{IceCube_2013,IceCube_2013_2,IceCube_2014}. The all-sky search over a period of 
$\sim988$~days resulted in 37 events in the 
energy range between $\sim30$~TeV and $\sim2$~PeV. The possibility of these neutrinos having a purely atmospheric origin is currently ruled out at $\sim5.7\sigma$. 
Whether their origin is Galactic or extragalactic remains unknown with several {\it ordinary} astrophysical sources being considered so far (for an excellent review see \cite{Cosmic_PeV}). 
An intriguing possibility related to new physics is that of PeV dark matter decay or annihilation. A {\it smoking gun} dark-matter-induced monochromatic neutrino line might be consistent with both, an apparent drop-off feature above PeV energies in the neutrino spectrum, and the fact that the three highest neutrino events have similar energies ($1041^{+132}_{-144}$~TeV, $1141^{+143}_{-133}$~TeV and $2004^{+236}_{-262}$~TeV). The case of dark matter decay has been considered in detail elsewhere \cite{Feldstein_13,Esmaili_13,Bai_13,Bat_14} but dark matter annihilation has been discarded for the following reason:

The rate of monochromatic neutrinos of energy $E_{\nu}$ produced by dark matter annihilation arriving at a detector on Earth of fiducial volume $V$ ($\sim1$~km$^3$ for IceCube) and nucleon number density $n_N$ 
($\sim5\times10^{23}$cm$^{-3}$, the number density of ice) has been estimated as \cite{Feldstein_13}:
\begin{equation}\label{base}
	\Gamma_{\rm Events}\sim V L_{\rm MW} n_N \sigma_N \left(\frac{\rho_\chi(R_\odot)}{m_\chi}\right)^2\left<\sigma_{\rm ann} v\right>\sim0.013~{\rm yr}^{-1},
\end{equation}
using a neutrino-nucleon scattering cross section $\sigma_N\sim9\times10^{-34}$cm$^2$ at $E_\nu=m_\chi=1.2$~PeV \cite{Gandhi_98}. Dark matter is assumed to annihilate homogeneously 
across the characteristic length of the Milky Way (MW) galaxy $L_{\rm MW}\sim 10$~kpc, having a density equal to the estimated local value $\rho_\chi(R_\odot=8.5~{\rm kpc})=0.4$~GeVcm$^{-3}$ (consistent with current estimates, see e.g. \cite{Bovy_12}) and an annihilation cross section {\it exclusively into neutrinos} and saturated at the {\it local} unitarity limit:
\begin{equation} \label{uni_local}	
	\left<\sigma_{\rm ann} v\right>\equiv\frac{4\pi}{m_\chi^2\beta_{\rm loc}},
\end{equation}
where $\beta_{\rm loc}\equiv v_{\rm loc}/c\sim10^{-3}$ is the typical {\it local} relative velocity of dark matter 
particles. 

With the estimate in Eq.~(\ref{base}), it seems that annihilation cannot account for the observed number of events. However, a more detailed calculation of the neutrino flux coming from {\it all} our MW halo should consider the following: (i) the change in dark matter density along the line of sight due to the radial dependence of the smooth dark matter distribution, which is enhanced towards the Galactic Centre; (ii) the contribution from dark matter substructure, and, more importantly, (iii) the unitarity limit depends on the relative velocity between dark matter particles at a given position along the
line of sight. Thus, in principle, without violating the unitarity bound, the annihilation cross section could be much larger in the cold substructures present in our halo than at the solar circle as assumed in Eq.~\ref{base}. This type of behavior is natural in Sommerfeld-enhanced models, where $(\sigma_{\rm ann} v)\propto1/\beta$ is a common feature due the presence of a new mediator acting between the annihilating particles (e.g. \cite{Hisano_04,Arkani_09,Lattanzi_09}). 

In this paper we consider in detail (i)-(iii) to compute the rate of neutrino events potentially observable by IceCube and produced by PeV dark matter annihilation in our Galactic halo. The paper is organized as follows: In Section \ref{sec_one}, we describe how we estimate the contributions from the smooth dark matter distribution and from substructure (see also Appendix). We study the cases of a constant $(\sigma_{\rm ann} v)$ and one where $(\sigma_{\rm ann} v)\propto1/\beta$. The expected neutrino rate is presented in Section \ref{sec_results} as well as some indications of the compatibility of the spatial event distribution in the sky with that expected in a dark matter annihilation scenario. In Section \ref{sec_four}, the possible origin of PeV dark matter particles and their associated minimum self-bound halo mass are discussed. Finally we present a summary and our conclusions in Section \ref{conclusions}.

\section{Galactic neutrinos from dark matter annihilation} \label{sec_one}

To estimate the enhancement over Eq.~(\ref{base}) due to the factors (i)-(iii) described above, we define the quantity $J(\Psi)$ (proportional to the so-called $J-factor$), as the line of sight integration of the dark matter density squared normalized to the product $\rho_\chi(R_{\odot})^2 L_{\rm MW}$ assumed in Eq.~(\ref{base}). 

\subsection{Smooth dark matter halo}\label{sec_smooth}

In the case of the smooth dark matter distribution we have\footnote{We fill follow closely the notation used in \cite{Yuksel_07}.}:
\begin{equation}\label{Jfactor_main}
	J_{\rm smooth}(\Psi)=\frac{1}{\rho_\chi(R_{\odot})^2 L_{\rm MW}}\int_0^{\lambda_{\rm max}}\rho_\chi^2\left(r=(R_\odot^2-2\lambda R_\odot {\rm cos}\Psi + \lambda^2)^{1/2}\right)d\lambda,
\end{equation}
where $\Psi$ is the angle relative to the Galactic centre, $\rho_\chi(r)$ is the radial density profile of the smooth component, and:
\begin{equation}
	\lambda_{\rm max} = R_\odot{\rm cos}\Psi + \left( R_{\rm 200}^2 - {\rm sin}^2\Psi R_\odot^2 \right)^{1/2},
\end{equation}
which truncates the integral at the virial radius of the halo, chosen as the radius where the average dark matter density is 200 times the critical density ($R_{\rm 200}$). 
To compute Eq.~(\ref{Jfactor_main}), we use an Einasto profile:
\begin{equation}\label{einasto}
\rho_\chi(r)=\rho_{-2}{\rm exp}\left(\frac{-2}{\alpha_e}\left[\left(\frac{r}{r_{-2}}\right)^{\alpha_e}-1\right]\right),
\end{equation}
where $\rho_{-2}$ and $r_{-2}$ are the density and radius at the point where the logarithmic density slope is $-2$, and $\alpha_e$ is the Einasto
shape parameter. We use the values of these parameters from the fit to the highest resolution level of the MW-size halo simulations from the Aquarius project (Aq-A-1 in \cite{Springel_08,Navarro_10}): $\alpha_e=0.17$, $\rho_{-2}=4\times10^{6}~{\rm M}_\odot {\rm kpc}^{-3}$, $r_{-2}=15.14~{\rm kpc}$, $M_{200}=1.84\times10^{12}$M$_\odot$, and $R_{200}=246$~kpc. We further normalize this profile to the assumed local dark matter density $\rho_\chi(r=R_\odot)=0.4$~GeVcm$^{-3}$. This value is within the range of current observational estimates, e.g. $\rho_\chi(R_\odot)=0.3~\pm~$0.1~GeVcm$^{-3}$ \cite{Bovy_2012}.

\subsubsection{Sommerfeld enhancement}\label{sec_SE_main}

Although the velocity dependence of $(\sigma_{\rm ann} v)$ in specific Sommerfeld-enhanced models is more complicated than a simple $1/\beta$ scaling, we will assume such behavior for simplicity noting that a particular model would not differ qualitatively from our main conclusions. 

We further assume that the dark matter particles have a Maxwellian velocity distribution without truncation. With such an assumption the $(\sigma_{\rm ann} v)\propto1/\beta$ scaling translates into an {\it average} that scales as $\left<\sigma_{\rm ann} v\right>\propto\left<1/\beta\right>=1/\sqrt{\pi}\sigma_{\rm vel}$, where $\sigma_{\rm vel}$ is the 1D velocity dispersion of dark matter particles in units of the speed of light. In this case, Eq.~(\ref{Jfactor_main}) is modified by replacing:
\begin{equation}\label{Jfactor_main_enh}
	\rho_\chi^2\left(\lambda\right)\rightarrow\left(\frac{1}{\sqrt{\pi}\sigma_{\rm vel}(\lambda)}\right)\rho_\chi^2\left(\lambda\right).
\end{equation}
We take the velocity dispersion profile for the MW halo as given by the spherically averaged coarse-grained pseudo phase space density $Q$:
\begin{equation}
	Q(r)\equiv\frac{\rho_\chi(r)}{\sigma_{\rm vel}^3(r)}\propto r^{\chi},
\end{equation}
where $\chi\sim-1.9$ and we normalize this relation to match $\sigma_{\rm vel}(r_{\rm max})\sim117$~km/s for the Aq-A-1 halo (see Tables 1 and 2 of \cite{Navarro_10}). 

In principle, a self-consistent approach would consider the actual velocity distribution associated to the Einasto profile, truncated to the local escape velocity, instead of a Maxwell Boltzmann distribution. Since a truncation does not significantly impact 
the integral that is needed to make the $\left<\sigma_{\rm ann} v\right>$ average, the scaling  $\left<\sigma_{\rm ann} v\right>\propto1/\sigma_{\rm vel}$ is thus preserved (see e.g. \cite{Bovy_09}). Although the specific shape of the velocity distribution derived self-consistently would deviate from a Maxwellian, the average $\left<\sigma_{\rm ann} v\right>$ is not overly sensitive to the precise shape of the distribution. As long as the Maxwellian assumption considers the accurate radial variation of the velocity dispersion (such as in our case), the final result will be approximately correct (see Figure 5 of \cite{Zavala_13a} and also Table I and Section 6 of \cite{Ferrer_13}). In any case, the signal we are exploring in this work is dominated by substructures, whose contribution is estimated with a model that does not assumes a specific velocity distribution (see Section \ref{sub_sec}). We therefore consider that the current approach to estimate the smooth dark matter contribution is sufficient for the purposes of this analysis. 

In physical models that have the Sommerfeld mechanism, the enhancement eventually saturates at low velocities due to the finite range of the interaction acting between the annihilating pair\footnote{Related to the non-zero mass of the force mediator, e.g. a Yukawa-like interaction.}. We take the saturated enhancement $S_{\rm max}$, relative to the value of the cross section at the solar circle (Eq.~\ref{uni_local}), as a free parameter.

\subsection{Dark matter substructure}\label{sub_sec}

To compute the contribution from substructure we use our new method based on a novel measure of dark matter clustering in phase space: Particle Phase Space Average Density ($P^2SAD$) \cite{Zavala_13a,Zavala_13b}. The method is calibrated to the Aquarius simulations, and uses a physically motivated model based on the stable clustering hypothesis \cite{Davis_77,Afshordi_10}, the spherical collapse model and tidal disruption of subhalos, to predict the behavior of $P^2SAD$ for masses below the resolution of current simulations. 
This prediction can then be used to compute signals that are sensitive to the small scale structure of dark matter such as annihilation.
The physical basis of our model gives it an advantage over most models that rely on simple extrapolations of the abundance, radial distribution and internal structure of subhalos, based on the behavior in the resolved regime. For completeness, we nevertheless present our results using both $P^2SAD$ and a current subhalo model (see Appendix \ref{sec_app} for details of the latter). 

The calibration of $P^2SAD$ with the Aquarius simulation is  presented in \cite{Zavala_13a}, while the methodology to compute the substructure contribution to dark matter annihilation is described in \cite{Zavala_13b}. A public version of a code that illustrates the use of $P^2SAD$ and our model is available online at \url{http://spaces.perimeterinstitute.ca/p2sad/}. 
In the following we simply give the equation that we need for the purposes of this paper.

The local subhalo boost $B(r)$ to the smooth dark matter annihilation rate is given by (see Eq. 9 of \cite{Zavala_13b}):
\begin{equation}\label{boost}
  B(r)\sim \frac{ \int d^3{\bf v} (\sigma_{\rm ann} v) \lim_{\Delta x \to 0} \Xi^{\rm subs}(\Delta x, v)}
  {\rho_\chi(r)\langle\sigma_{\rm ann} v\rangle_{\rm smooth}},
\end{equation}
where $\Xi^{\rm subs}$, a function of both relative velocity and separation ($\Delta x$) between the annihilating pair of particles, is equivalent to $P^2SAD$ in the regime dominated by substructures (small separations in phase space)\footnote{$P^2SAD$ is a full measure of the clustering of dark matter in phase space, and thus, it does not distinguish the smooth from the substructure contribution. However, at small scales (i.e. small separations in phase space), $P^2SAD$ is dominated by substructures (see Eq. 7 of \cite{Zavala_13b} for the numerical value of $P^2SAD$ where this transition occurs in a MW halo).}. The quantities in the denominator in Eq.~(\ref{boost}) correspond to the smooth dark matter distribution. The physically motivated model that we present in \cite{Zavala_13b} can then be used to predict $\lim_{\Delta x \to 0} \Xi^{\rm subs}(\Delta x, v)$ for any minimum subhalo mass. {\it Since substructure is naturally embedded in $P^2SAD$, our model does not require an assumption about the velocity distribution function of dark matter}. Thus, any velocity dependence of $(\sigma_{\rm ann} v)$ can be easily accommodated by performing the simple integral in Eq.~(\ref{boost}). The particular $1/\beta$ scaling of the Sommerfeld-enhanced models we consider here is
therefore straightforward\footnote{See Eqs. 32 and 35 for the specific equations that were used in our calculations for the cases of $(\sigma_{\rm ann} v)=$cte and $(\sigma_{\rm ann} v)\propto1/\beta$, respectively.}.

The $J-factor$ from substructures is then simply given by:
\begin{equation}\label{Jfactor_subs_p2sad}
	J_{\rm subs}(\Psi)=\frac{1}{\rho_\chi(R_{\odot})^2 L_{\rm MW}}\int_0^{\lambda_{\rm max}}B(\lambda)\rho_\chi^2\left(\lambda\right)d\lambda.
\end{equation}
We note that the Aquarius simulations were done in the context of a WMAP1 cosmology whose parameters are different from those currently preferred. In particular, $\sigma_8$ (the rms amplitude of linear mass fluctuations in $8~h^{-1}$~Mpc spheres at redshift zero) is lower and $\Omega_m$ is higher in the latter. This produces compensating effects in the abundance and clustering of dark matter haloes. Nevertheless, due to the chosen cosmology, there might be a small overestimate of the abundance and central densities of subhalos, i.e., an overestimate of the substructure contribution to the annihilation rate. We also note that Aq-A-1 is only a particular realization of a halo with similar global properties to that of our Galactic halo; it is not expected however to be a detailed match. Observational uncertainties allow for a broad range of density profiles, which
impacts the predicted signal of the smooth halo (see the end of Section III). Surprisingly, the contribution from substructures might be relatively insensitive to different MW realizations, as long as the MW halo mass
is not too far from that of the Aq-A-1 halo\footnote{Current estimates of the MW halo virial mass cover the range $\sim1.0-2.0\times10^{12}$~M$_\odot$ (e.g. see Section 5.1 of \cite{BK_2012})}. This is a conclusion
based on the analysis of $P^2SAD$ across the different simulations of the Aquarius project \cite{Zavala_13a}.

\section{Results}\label{sec_results}

If we take the average of $J(\Psi)$ over the solid angle $\Delta\Omega=2\pi(1-\rm{cos}\Psi)$:
\begin{equation}\label{J_ave}
	J_{\rm \Delta\Omega}(\Psi)=\frac{2\pi}{\Delta\Omega}\int_{{\rm cos}\Psi}^1J(\Psi')d({\rm cos}\Psi'),
\end{equation}
then we can finally estimate the average number of PeV neutrinos ($E_\nu=m_\chi=1.2$~PeV), with a dark matter annihilation origin, expected in IceCube within an angle $\Psi$  from the Galactic Centre (visible fraction $\Delta\Omega/4\pi$ of the whole sky):
\begin{equation}\label{events_eq}
	\Gamma_{\rm Events}(\Psi)= V L_{\rm MW} n_N \sigma_N \left(\frac{\rho_\chi(R_\odot)}{m_\chi}\right)^2\left<\sigma_{\rm ann} v\right>_{\rm loc} \left(\frac{\Delta\Omega}{4\pi}\right)J_{\rm \Delta\Omega}(\Psi),
\end{equation}
where the {\it local} annihilation cross section is at the {\it local} unitarity limit. We note that, following our definitions, we take $\beta_{\rm loc}=\sigma_{\rm vel, loc}\sim3.7\times10^{-4}$ instead of $\beta_{\rm loc}=10^{-3}$ as used in Eq.~(\ref{base}).

\begin{figure}
\centering
\subfigure{\includegraphics[width=0.49\textwidth]{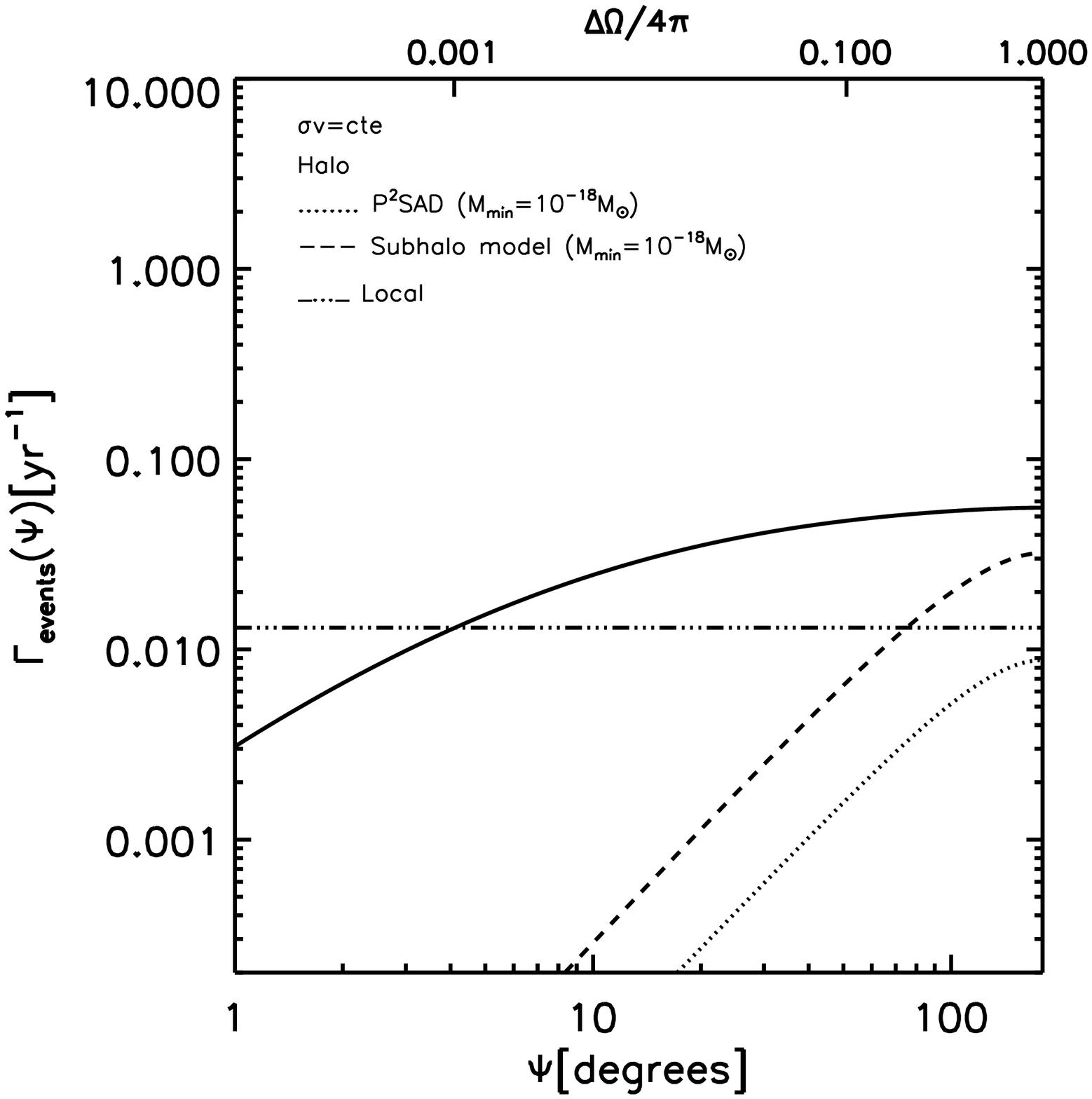}}
\subfigure{\includegraphics[width=0.49\textwidth]{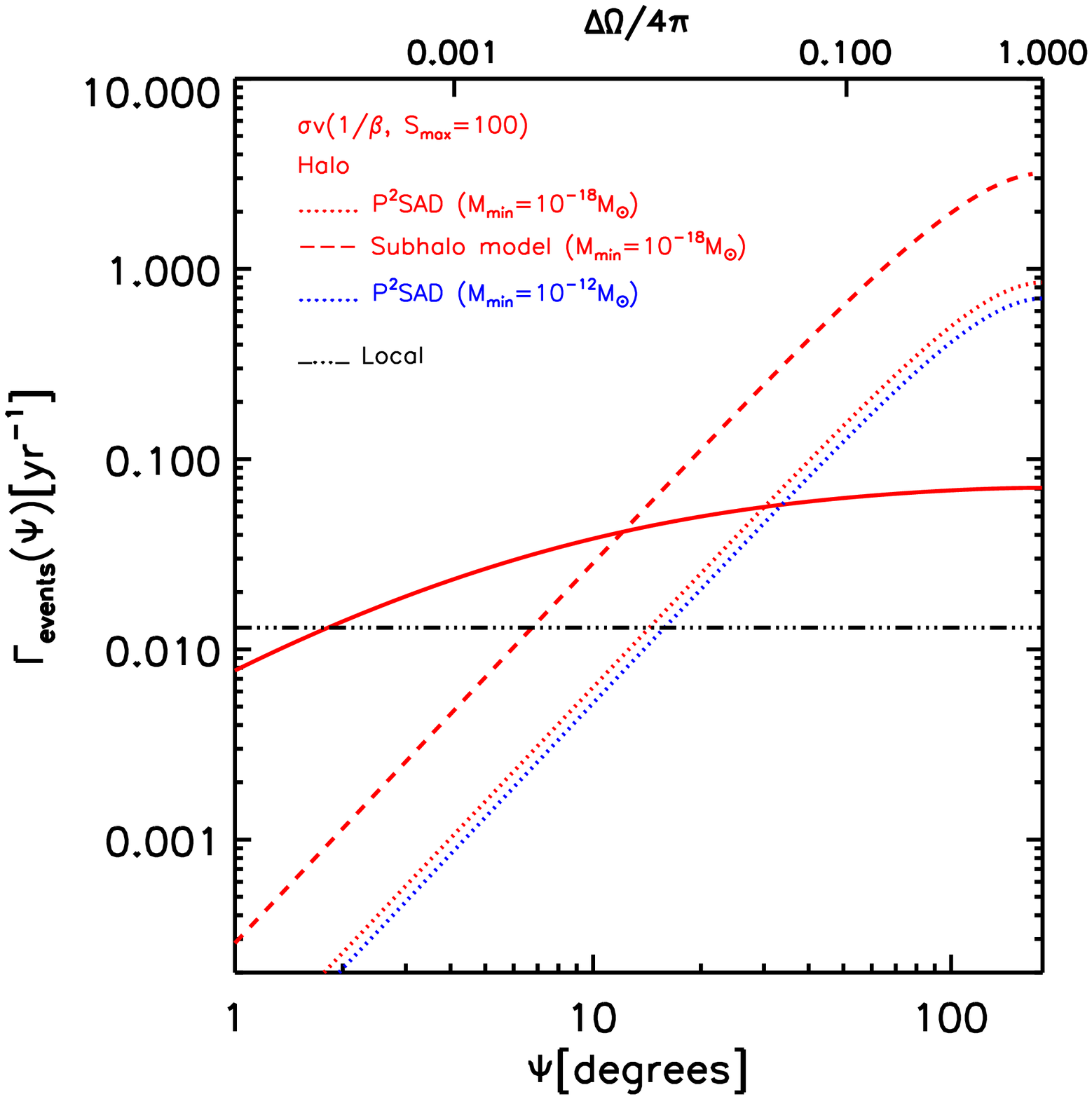}}
\caption{Average rate of neutrinos ($E_\nu=m_\chi=1.2$~PeV) expected in IceCube within an angle $\Psi$ from the Galactic Centre (solid angle $\Delta\Omega$) produced by dark matter annihilating {\it exclusively} into neutrinos along the line of sight. The solid lines are for the smooth dark matter distribution while the dashed and dotted lines are the contributions from substructure down to a mass of $10^{-18}$M$_\odot$ using two different models to account for substructure. For the left panel, we assume a constant $\left<\sigma_{\rm ann} v\right>$ equal to the
unitarity limit set {\it locally} (Eq.~\ref{uni_local} with $\beta_{\rm loc}\sim3.7\times10^{-4}$). For the right panel (red lines), $\left<\sigma_{\rm ann} v\right>$ is normalized to the same local value but it scales as $1/\beta$ (Sommerfeld enhancement) until it saturates at $S_{\rm max}=100$ times the local value. The blue dotted line is for $m_{\rm min}=10^{-12}$M$_\odot$. The dot-dashed line in both panels is the estimate according to Eq.~(\ref{base}).}
\label{Jfactor}
\end{figure}

The results of our calculation are shown in Fig.~\ref{Jfactor} for the cases where $\left<\sigma_{\rm ann} v\right>=\left<\sigma_{\rm ann} v\right>_{\rm loc}={\rm cte}$ (left panel) and in the case where $(\sigma_{\rm ann} v)\propto1/\beta$ with a saturation at $S_{\rm max}=100$ (right panel). The solid lines are for the smooth dark matter distribution while the dashed lines show the contribution from substructure all the way down to $m_{\rm min}=10^{-18}$M$_\odot$ (a representative free streaming mass for PeV dark matter particles, see section \ref{free_sec} below). We present the contribution from substructure using two models: the $P^2SAD$ approach (dotted lines, see Section \ref{sub_sec}) and the subhalo model (dashed lines, see Appendix). They both agree in the spatial distribution of subhalos, but the latter is a factor of $\sim3$ larger than the former. We note that the fact that both models agree reasonably well despite their fundamentally different approach is because: (i) they are both calibrated to the same $N-$body simulation (Aquarius) and (ii) they both take into account the flattening  of the dark matter power spectrum at small scales. Nevertheless, there is a clear difference in normalization that can be traced back to the inaccuracy of the subhalo model in describing $P^2SAD$ {\it in the resolved regime}, overestimating it by a factor of $2-3$ (see Fig. 5 of \cite{Zavala_13a} and discussion therein). The reason for this is a combination of the simplifications made in the subhalo model, perhaps more importantly: (i) the need to assume a velocity distribution function for the subhalos (a Maxwellian), and (ii) the use of an average relation for the concentration-mass subhalo relation, instead of considering the full  {\it radially-dependent} concentration distribution as a function of subhalo mass (subhalo concentrations rise towards halo centres, see e.g. \cite{Springel_08}). This highlights the importance of $P^2SAD$ as a quantity that can be used directly to estimate signals that are sensitive to the
small scale structure of dark matter, such as dark matter annihilation.

The total rate of neutrinos arriving at IceCube over the full sky can be read off from Fig.~\ref{Jfactor} by taking $\Gamma_{\rm Events}(\Psi=180^\circ)$. For the constant  $(\sigma_{\rm ann} v)$ case, the effect of substructure in the total annihilation rate is less but of the order of the smooth distribution (roughly consistent with the low end of previous studies, e.g., \cite{Springel_08_2,Fornasa_13}). Together, both contributions add up to $\Gamma_{\rm Events}\sim0.06-0.09$\footnote{The lower value is from the estimate using $P^2SAD$, while the larger value comes from the subhalo model.}, still too low to account for the observed number of events but a factor of several larger than the simple estimate in Eq.~(\ref{base}).  For the Sommerfeld-enhanced case, the role of substructures becomes dominant and already for $S_{\rm max}=100$, the predicted number of events per year is $\sim0.92-3.1$, which is of the order of the observed rate of PeV neutrinos.

The effect of $m_{\rm min}$ in our results is relatively small due to the flattening of the dark matter power spectrum at small scales. For instance, increasing the minimum mass by 6 orders of magnitude only changes the subhalo contribution by a factor of $\sim1.4$ (blue dotted line in the right panel in Fig.~\ref{Jfactor}). Thus, although we discuss it in Section \ref{free_sec}, the precise value of the damping mass scale for PeV dark matter is not a major uncertainty.

From Fig.~\ref{Jfactor}, we can readily see that while the subhalo contribution might become dominant over large angles, at small angles the smooth distribution clearly dominates. Thus, a generic expectation for a signal with a dark matter origin (more relevant for annihilation but also for decay) is a larger number of neutrino events in the direction of the Galactic Centre. For instance in the Sommerfeld-enhanced case considered here, the transition between both regimes occurs at $\sim10-30^\circ$. This transition however, depends on the interplay between the normalization of the cross section, the value of $S_{\rm max}$ and, to lesser extent, the value of $m_{\rm min}$. For example, one can achieve the desired rate with a smoother distribution at all angles by increasing the value of $S_{\rm max}$ and reducing the normalization of the cross section. We note that we calculated
Fig. 1 assuming $E_\nu=m_\chi=1.2$~PeV, for higher neutrino energies, such as the case of the recently reported $\sim2$~PeV event \cite{IceCube_2014}, the normalization of the cross section would be lower by a factor of 
$(2/1.2)^2\sim2.8$. This can be compensated by increasing the value of $S_{\rm max}$ accordingly. Notice also that due to the low number PeV events, the measured event rate with error bars corresponding to $2\sigma$
errors (based on Poisson statistics \cite{Gehrels_1986}) is $1.1^{+2.2}_{-0.9}$~yr$^{-1}$, which leaves a freedom on the normalization of the predicted signal of $\mathcal{O}(10)$.

\begin{figure}
\centering
\includegraphics[height=10.0cm,width=10.0cm]{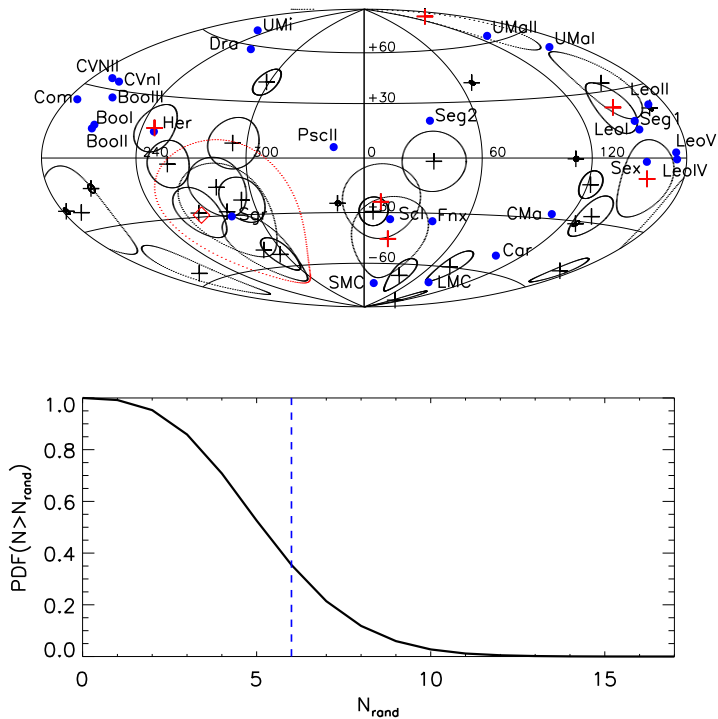}
\caption{Top panel: Aitoff projection in equatorial coordinates of 33 of the 37 high energy neutrino events reported by IceCube (crosses). We removed from the sample the two events with the largest angular errors ($\gtrsim42^\circ$), and also two more (numbers 28 and 32 in \cite{IceCube_2014}) that are very likely produced in cosmic ray air showers. The median angular error in the location of each event is shown with a circle. The blue circles mark the locations of 26 MW satellites. The six red crosses are neutrinos coincident with the position of at least one satellite (excepting Sagittarius). The red  square marks the position of SgrA$^\ast$. The red line circles all eight events in the inner halo region (defined arbitrarily as $40^\circ$ around Sagittarius). Bottom panel: Discrete cumulative probability distribution function of randomly drawing 25 points in the sky, excluding the inner halo region, and having $N>N_{\rm rand}$ neutrinos coincident with at least one of these 25 points (within the angular errors). The dashed blue line marks the 6 actual events coincident with 5 MW satellites. The probability that this number (or higher) occurs randomly is $\sim35\%$.}
\label{events}
\end{figure}

As is clear in Fig.~\ref{Jfactor}, in a Sommerfeld-enhanced model of annihilation, subhalos are more visible in the sky than if $(\sigma_{\rm ann} v)={\rm cte}$; the largest nearby subhalos might therefore appear as individual sources (for a sky map realization of dark matter annihilation in a Sommerfeld case see e.g \cite{Kuhlen_09}). 
In addition to the three $\sim$PeV neutrinos, the IceCube collaboration has also reported 34 neutrino events at lower energies ($0.03~{\rm PeV}<E_\nu<0.4$~PeV) in an all-sky search \cite{IceCube_2013_2}. These lower energy neutrinos could be the result of a continuum neutrino emission from annihilation, while the high energy events would be the result of monochromatic annihilation (see Section \ref{conclusions}). Even though the IceCube collaboration has performed a point source analysis of these events and found no strong evidence for spatial clustering, it is still interesting to investigate the compatibility of the observed all-sky map with a predicted dark matter signal. This was done for the case of dark matter decay in \cite{Bai_13}.  Although we do not attempt to investigate this rigorously in the case of dark matter annihilation, we note that the distribution has some intriguing features. In Fig.~\ref{events} we show the spatial distribution of neutrino events in equatorial coordinates (crosses). Each event is circled by the angular error in its position (based on Table 1 of \cite{IceCube_2013_2}). We have removed the events with the largest errors, $\sim43^\circ$ and $\sim46^\circ$\footnote{The center of this event lies near the Galactic Centre.}, and also two events that are almost certainly produced
in cosmic ray air showers (see Table I of \cite{IceCube_2014}).

Eight of the 33 events are near the Galactic Center (SgrA* marked with a red square), which in a dark matter interpretation would be associated with the smooth dark matter distribution. With blue circles we mark the locations of 26 MW satellites easily identified by the legends. It is interesting that 6 of the 25 events outside the inner halo region are compatible, within the angular errors, with coming from 5 MW satellites: Hercules, Sculptor, Segue 1, Sextans and Ursa Major II. Notice also that another 7 satellites are barely outside the reported angular errors of the events: Fornax, LMC, Leo I-II, Leo III-IV and Ursa Major I. 

To estimate how likely is that the coincidence between some of the neutrino events with some of the MW satellites occurs by chance, we randomly draw 25 locations from the sky (uniform distribution in right ascension and declination), excluding a $40^\circ$ region centered in the Sagittarius galaxy. This exclusion region was selected so that it roughly encompasses all eight events possibly associated with the inner halo. The discrete cumulative probability distribution of having $N>N_{\rm rand}$ events coincident with at least one of the 25 random locations is shown in the bottom panel of Fig.~\ref{events}. The probability of having 6 
or more random matches is $\sim35\%$. Although this high probability is certainly consistent with a random coincidence, 
the statistics are limited at this point and it would be interesting to study such association once more data is collected. We note that most of the events come from the southern hemisphere, which might be due to some neutrinos being absorbed as they pass through the Earth and reach the detector from below.

\begin{figure}
\centering
\subfigure{\includegraphics[width=0.49\textwidth]{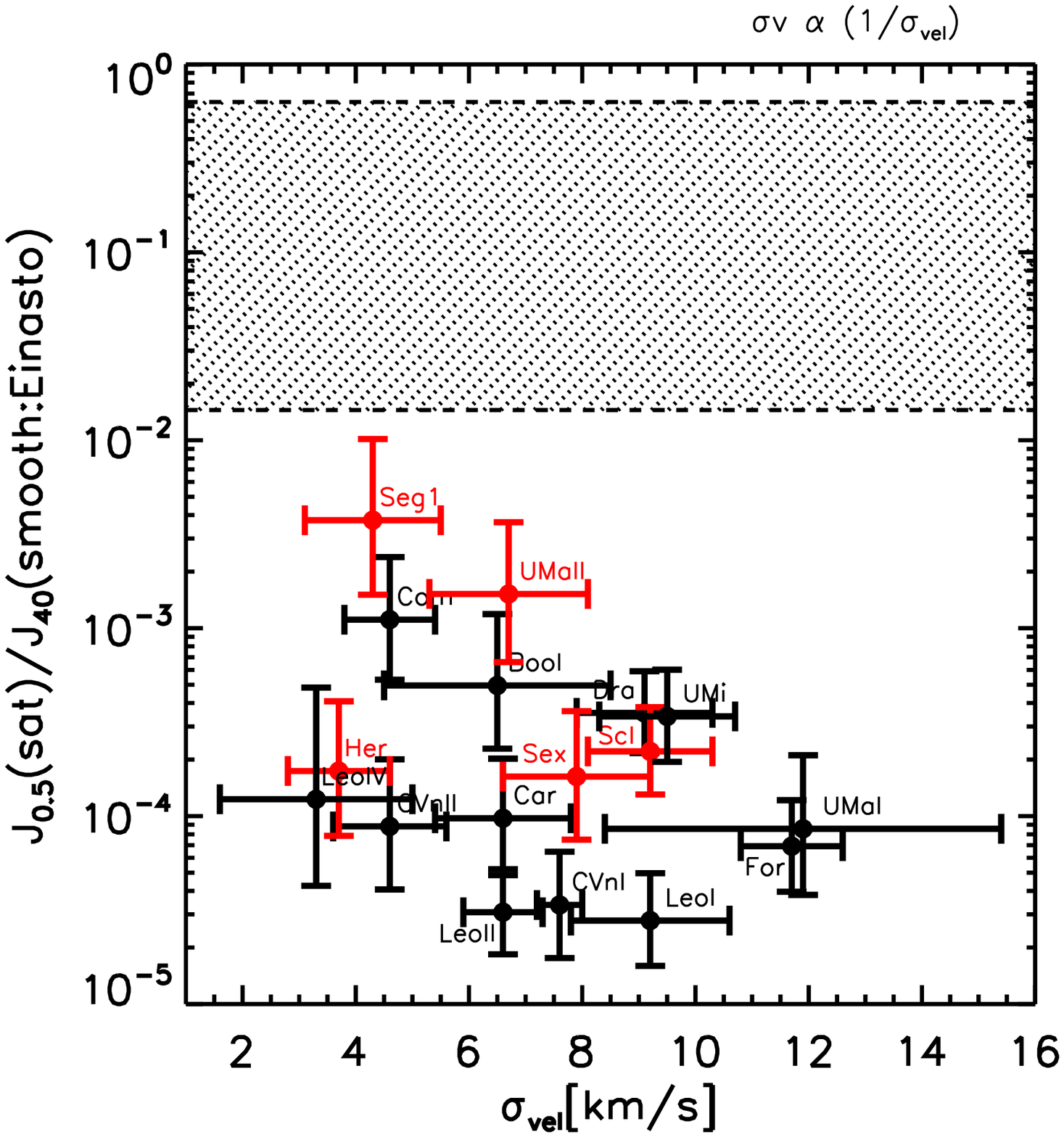}}
\subfigure{\includegraphics[width=0.49\textwidth]{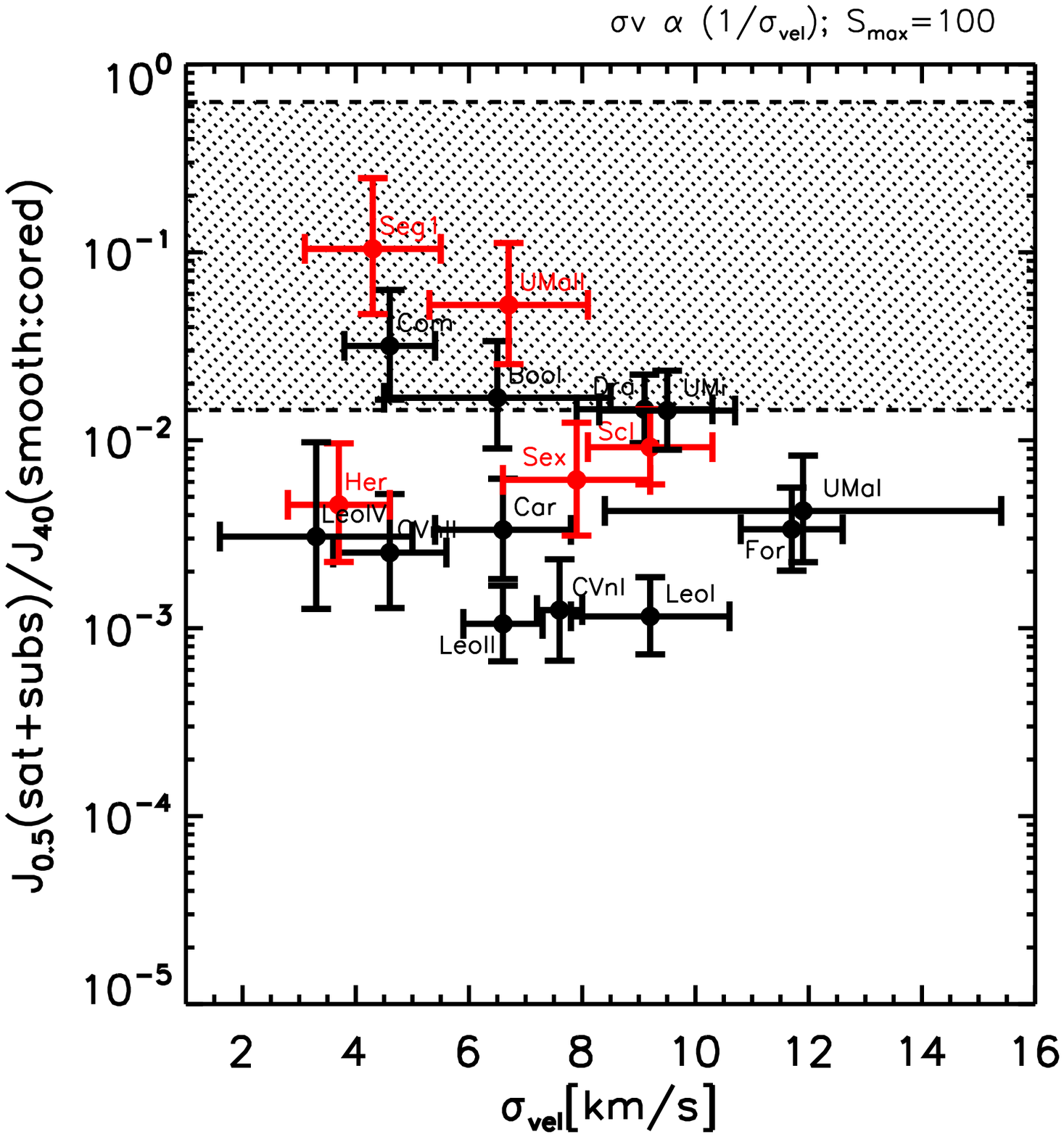}}
\caption{Ratio of the expected number of neutrino events within $0.5^{\circ}$ of the center of 17 MW satellites to that within $40^{\circ}$ from the Galactic Centre produced by dark matter annihilating {\it exclusively} into neutrinos. The $J-factor$ (Eq. \ref{J_ave}) 
for each galaxy was taken from \cite{Fermi_13} (assuming a NFW profile, see their Table I) scaled up by the {\it average} Sommerfeld enhancement given by the corresponding velocity dispersion (taken from \cite{Walker_2009}): 
$\left<\sigma_{\rm ann} v\right>\propto 1/{\bar \sigma_{\rm vel}}$. In the left panel, the MW halo is assumed to have an Einasto profile as described in Section \ref{sec_smooth} while in the
right a constant density core of $8.5$~kpc is assumed. In addition, the satellites have a sub-substructure boost with a Sommerfeld enhancement saturated at $S_{\rm max}=100$; see text for details. The horizontal dashed line marks the $1\sigma$ region of the hypothetically observed satellite:Galactic-Centre ratio of events, 1:8.}
\label{Jfactor_dwarfs}
\end{figure}

If dark matter annihilation in the MW satellites is indeed responsible for some of the cosmic neutrino events, it would be expected that those satellites with the largest predicted rates are precisely those that
have a matching event in Fig.~\ref{events}. To check this, we take the $J-factors$ estimated in \cite{Fermi_13} for the MW satellites that have enough kinematical information (17 out of the 26). They were computed 
using estimates of the dark matter mass distribution in each galaxy compiled from several references (see their Table I). The annihilation signal is computed within $0.5^{\circ}$ of the centre of each satellite, and we consider only the case where the dark matter density profile is assumed to be NFW  
\cite{NFW_97}\footnote{Although current kinematical observations of the stars in the satellites are not sufficient to unambiguously discriminate a NFW from a cored-like profile (e.g. Burkert), the difference in the $J-factor$ between both cases is less than a factor of $\sim2$ (see Table I of \cite{Fermi_13}).}. Since the $J-factor$ in each galaxy was computed assuming $\left<\sigma_{\rm ann} v\right>={\rm cte}$, we scale it up to account for a
Sommerfeld enhancement. Instead of a local boost as in Eq.~\ref{Jfactor_main_enh}, we boost the $J-factor$ given in \cite{Fermi_13} by a factor proportional to $1/{\bar \sigma_{\rm vel}}$, where 
${\bar \sigma_{\rm vel}}$ is the average velocity dispersion of each satellite (taken from \cite{Walker_2009}).  We then compute the ratio of each enhanced $J-factor$ to that of the smooth dark matter distribution in the
MW halo within $40^{\circ}$ of the Galactic Centre. The result is shown in the left panel of Fig.~\ref{Jfactor_dwarfs} as a function of the velocity dispersion of each satellite. Highlighted in red are the five systems that are coincident
with one neutrino event. Two of these are ranked the highest among all satellites, while the other three are among the next seven satellites with the largest $J-factors$. It is thus not unexpected that these five satellites are the ones giving a neutrino signal, although perhaps Coma Berenices, Bootes I, Draco and Ursa Minor, would also be expected to produce a signal. However, given the low number of events and the large error bars in the $J-factors$, no firm conclusion can be reached at this point.

The number of observed events within $40^{\circ}$ from the Galactic Centre is 8, since each satellite is coincident with one event (except for Sculptor), then, hypothetically, we have an {\it observed} 1: 8 satellite-Galactic-Centre ratio of events. Due to the low number of counts, the corresponding $1\sigma$ region around this ratio (based on Poisson statistics) is quite broad (dashed region in Fig.~\ref{Jfactor_dwarfs}). Still, the predicted signal from all
satellites lies below this region, i.e., there should be many more events around the Galactic Centre. A possibility is to reduce the annihilation rate in the center of the MW halo by modeling the density distribution with a cored instead of an Einasto profile. A large core of size $\sim8.5$~kpc is actually allowed by current data (see e.g. \cite{Bovy_2013,Salucci_2013}), and would decrease the number of events around $40^{\circ}$ from the Galactic Centre
by roughly an order of magnitude. Another uncertainty in the predicted satellite-Galactic-Centre ratio of events is that of the appropriate sub-substructure boost to the MW satellites, which was assumed to be zero in the left panel of 
Fig.~\ref{Jfactor_dwarfs}. This is motivated by the fact that dark matter subhalos are subjected to tidal forces that rapidly strip their most loosely bound material, this includes the abundant sub-substructure outside their tidal
radii. For instance, it has been estimated that the sub-substructure boost in a fraction of the MW satellites shown in Fig.~\ref{Jfactor_dwarfs} lies in the range: $12\%-31\%$ \cite{MASC_2011}. This is however in the case where
$\left<\sigma_{\rm ann} v\right>={\rm cte}$, if there is a Sommerfeld enhancement with $S_{\rm max}=100$, then, since the MW satellites we are considering have $\sigma_{\rm vel}$ of $\mathcal{O}(10~{\rm km~s}^{-1})$, their sub-subhalos would be all in the saturated regime. Thus, the sub-substructure boost would increase roughly by a factor $\sim S_{\rm max}/S_{\rm MW-sat}$ relative to the $\left<\sigma_{\rm ann} v\right>={\rm cte}$ case\footnote{Ideally, this full calculation would be done using $P^2SAD$, however, this cannot be done since $P^2SAD$ has been calibrated only at the scale of MW-size halos. A scale-dependence in Eq.~\ref{boost} cannot be discarded at present.}. If we consider then the case of a MW halo with a core and a Sommerfeld-enhanced sub-substructure boost, we obtain the right panel of Figure \ref{Jfactor_dwarfs}. This prediction is in much closer agreement to the hypothetically observed situation.

\section{PeV dark matter abundance and microhalos}\label{sec_four}

\subsection{Thermal relic abundance}\label{thermal}

Thermal production of very massive particles is dismissed as an explanation for the observed abundance of dark matter since
the unitarity bound in the early universe would imply an abundance today that overcloses the Universe \cite{Griest_90}. For a constant $\left<\sigma_{\rm ann} v\right>$ set at freeze-out, \cite{Griest_90} estimated an upper bound to the dark matter mass $m_\chi<340$~TeV, for a Majorana fermion in order to have $\Omega_\chi h^2\lesssim1$. However, for the case where the cross section is enhanced by a Sommerfeld mechanism, annihilation proceeds beyond freeze-out, reducing the relic abundance (e.g. \cite{Zavala_10_2,Dent_10}) making the unitarity mass bound weaker. This was already partially estimated by \cite{Griest_90} finding that $m_\chi<550$~TeV 
($\Omega_\chi h^2\sim1$) for the case we have studied here, i.e., considering the unitarity limit and assuming $(\sigma_{\rm ann} v)\propto1/\beta$ across freeze-out and beyond. The impact of the kinetic decoupling temperature was however not considered in this calculation. After kinetic decoupling, the temperature of the dark matter particles drops as $a^{-2}$, while the temperature of radiation drops as $a^{-1}$ (where $a$ is the scale factor). The reduced dark matter velocities imply a larger boost to the annihilation cross section reducing the relic abundance substantially. In the case where $(\sigma_{\rm ann} v)\propto1/\beta$, the relic density decays logarithmically. We follow closely \cite{Dent_10}, where the relic abundance for the latter case was computed in detail. 

For constant s-wave annihilation, $\left<\sigma_{\rm ann} v\right>_0={\rm cte}$, the observed abundance of a thermal relic today is given by:
\begin{equation}
	\Omega_\chi h^2 \sim 2.757\times10^8 \left(\frac{m_\chi}{{\rm GeV}}\right)Y_\infty,
\end{equation}
where:
\begin{equation}
	Y_\infty=\frac{3.79~x_f}{(g_{\ast,S}/g_\ast^{1/2})M_{\rm Pl}m_\chi\left<\sigma_{\rm ann} v\right>_0},
\end{equation}
where $x_f=m_\chi/T_f$ establishes the freeze-out temperature $T_f$, $M_{\rm Pl}$ is the Planck mass, and $g_\ast$ ($g_{\ast,S}$) are the effective degrees of freedom for the total energy (entropy) density of the Universe; $g\ast\neq g_{\ast,S}$ only if there are relativistic particles that are not in equilibrium with the photons. In the Standard Model this only happens at temperatures lower than neutrino decoupling $T<2-3$~MeV (e.g. \cite{Steigman_12}). Since we are always in a regime with larger temperatures we will take $g\ast=g_{\ast,S}$. 

In the Sommerfeld-enhanced case, we can write the cross section in terms of $x=m_\chi/T$: $\left<\sigma_{\rm ann} v\right>=\left<\sigma_{\rm ann} v\right>_0 x^{1/2}$, where 
$\left<\sigma_{\rm ann} v\right>_0=4\pi/(m_\chi^2\sqrt{\pi})\sim8.28\times10^{-29}$cm$^3$/s. The relic density in this case is reduced by a factor of (combining Eqs. 17 and 22 of \cite{Dent_10}):
\begin{equation}\label{supr}
	\frac{Y_\infty^{SE}}{Y_\infty}=\frac{1}{2}\frac{x_{f,SE}^{1/2}}{x_f}\frac{(T_f/T_{\rm kd})^{1/2}}{(T_f/T_{\rm kd})^{1/2}-1+1/2{\rm ln}\left(T_{\rm kd}/T_{\rm sat}\right)},
\end{equation}
where $T_{\rm kd}$ and $T_{\rm sat}$ are the radiation temperatures at kinetic decoupling and at the moment where the Sommerfeld enhancement finally saturates, respectively; $x_{f, SE}$ gives the freeze-out temperature in the Sommerfeld-enhanced case. The latter and $x_f$ are computed using the following formula, with $n=-1/2$ and $n=0$, respectively:
\begin{equation}
	x_{f}(n)={\rm ln}\left[(n+1)a_\chi\lambda_\chi\right]-(n+1/2){\rm ln}\left[{\rm ln}\left[(n+1)a_\chi\lambda_\chi\right]\right],
\end{equation} 
where $a_\chi=0.145(g/g_\ast)$ ($g$ is the number of degrees of freedom of the dark matter particle; $g=2$ for a Majorana fermion) and 
$\lambda_\chi=\sqrt{\pi/45}(g_{\ast,S}/g_\ast^{1/2})M_{\rm Pl}m_\chi\left<\sigma_{\rm ann} v\right>_0$. The radiation temperature at saturation, $T_{\rm sat}<T_{\rm kd}$, is directly related to the saturation velocity through the temperature of the dark matter particles after kinetic decoupling $T_\chi=T^2/T_{\rm kd}$. The largest suppression possible to the relic density occurs when $T_{\rm kd}=T_f$. In this limit, $T_{\rm sat}=\sqrt{T_{\rm \chi, sat} T_f}$, thus:
\begin{equation}
	\frac{T_{\rm kd}}{T_{\rm sat}}=\sqrt{\frac{T_f}{T_{\rm \chi, sat}}}\sim\frac{1}{\sigma_{\rm vel}({\rm sat})}\equiv S_{\rm sat},
\end{equation}
where we have assumed that at freeze-out, although the dark matter particles are already non-relativistic, their velocities are still very large. For $m_\chi=1$~PeV, $x_{f}\sim25.5$ and $x_{f,SE}\sim26.5$. Thus the largest suppression in Eq.~(\ref{supr}) for PeV particles is given by:
\begin{equation}
	\frac{Y_\infty^{SE}}{Y_\infty}\sim0.2\frac{1}{{\rm ln}(S_{\rm sat})}.
\end{equation}
In the example we have considered in Fig.~\ref{Jfactor}, $S_{\rm max}=100$ corresponding to $\sigma_{\rm vel}~c\sim1{\rm km/s}$; we can then estimate the relic density to be:
\begin{equation}\label{abundance}
	\Omega_\chi h^2 (m_\chi=1{\rm PeV}, T_{\rm kd}=T_f, S_{\rm max}=100)\sim0.47
\end{equation}
which is clearly inconsistent with the observed dark matter abundance. Since $\Omega_\chi$ depends only logarithmically on $S_{\rm sat}$, an extremely large saturation value would be needed to reduce the relic abundance to observed values. 

In the previous calculation we have assumed the particle content of the Standard Model (i.e., $g_\ast(T_{\rm kd}=T_f)=107$), but the freeze-out temperature is very large for PeV dark matter, $T_f\sim4\times10^4$~GeV. At these temperatures, the questions of how many extra degrees of freedom there are and when do they decouple remain open. Increasing this number would reduce the relic abundance, but since the dependence on $g_\ast$ is not strong, a substantial change would be required to reduce the abundance to the observed value.

An alternative is to consider a more radical departure from the standard thermal relic calculation. In particular, the assumption of a purely radiation dominated Universe might be broken if, for example, there exists new unstable massive particles with couplings too weak too maintain thermal equilibrium (for a description of this possibility see \cite{Watson_10,Hooper_13} and references therein). These particles would then naturally dominate the energy density of the Universe. If they decay into relativistic particles and reheat the Universe to a temperature below the freeze-out temperature ($T_{\rm RH}<T_f$), then the abundance of dark matter particles would be reduced by a factor of 
$\left(T_{\rm RH}/T_f\right)^3$. Thus, given Eq.~(\ref{abundance}), the reheating temperature has to be just slightly lower than $T_f$ to get the correct relic abundance, $T_{\rm RH}\sim 0.62~T_f$.

\subsection{Minimal halo mass}\label{free_sec}

In a PeV dark matter scenario, the hierarchy of dark matter self-bound structures will extend to much lower masses than in the case of standard Weakly Interactive Massive Particles (WIMPs). The comoving free streaming length is roughly given by the time $t_{\rm nr}$ when the particles become non-relativistic: $R_{\rm fs}\sim 2 ct_{\rm nr}/a_{\rm nr}\propto1/m_{\chi}$. For massive particles $t_{\rm nr}$ occurs in the radiation dominated epoch. Since the typical WIMP masses are of $\mathcal{O}(100{\rm GeV})$ with free streaming masses of $\mathcal{O}(10^{-6}{\rm M}_\odot)$, then for PeV dark matter particles, $M_{\rm fs}\sim 10^{-18}$M$_\odot$. 

The actual damping scale for PeV dark matter will depend on its interactions with Standard Model particles. If the dark matter particles are produced thermally, then their coupling to the thermal bath would erase any fluctuations until chemical decoupling (freeze-out). Afterwards, elastic scattering between dark matter and Standard Model particles would still damp fluctuations until they finally decouple kinetically and free stream. The
final damping scale would then depend on the kinetic decoupling temperature (e.g. \cite{Bringmann_09}):
\begin{equation}
	M_{\rm fs}=2.9\times10^{-6}\left(\frac{1+{\rm ln}(g_\ast^{1/4}T_{\rm kd}/50{\rm MeV})/19.1}{(m_\chi/100{\rm GeV})^{1/2}g_\ast^{1/4}(T_{\rm kd}/50{\rm MeV})^{1/2}}\right)^3M_\odot,
\end{equation}
where $g_\ast$ is evaluated at $T=T_{\rm kd}$. Taking the assumptions from section \ref{thermal} above, $T_{\rm kd}\sim T_{\rm f}\sim4\times10^4$~GeV, then $M_{\rm fs}\sim10^{-21}$M$_\odot$. The value of the kinetic decoupling temperature  is however model dependent and if $T_{\rm kd}\ll T_f$, then $M_{\rm fs}$ would be much higher than this. We have taken the simple estimate, $M_{\rm fs}=10^{-18}$M$_\odot$ mentioned in the previous paragraph as a benchmark value. As we discussed in section \ref{sec_results}, the estimated neutrino rate is not very sensitive to the precise value of $M_{\rm fs}$.

\section{Discussion and Conclusions}\label{conclusions}

The announcement by the IceCube collaboration of the detection of over thirty neutrino events with a likely cosmic origin has been received with excitement raising a significant interest in discovering the responsible sources. 
At this moment, {\it ordinary} astrophysical sources (Galactic and/or extragalactic) could be responsible for the signal (for a review see \cite{Cosmic_PeV}), but the possibility of a dark matter origin is intriguing due to the connection with new physics beyond the Standard Model. 

The decay of PeV dark matter  particles into neutrinos has been proposed recently \cite{Feldstein_13,Esmaili_13,Bai_13,Bat_14} while the case of PeV annihilating particles was dismissed invoking the unitarity bound to the annihilation cross section. In this work, we have revised the latter claim and compute in greater detail the expected rate of monochromatic neutrinos from PeV dark matter annihilation. We find that the unitarity limit can be satisfied and still produce sufficient PeV neutrinos if the cross section is enhanced by a Sommerfeld mechanism. In the simple case of $(\sigma_{\rm ann} v)\propto1/v$, the unitarity bound allows for a larger annihilation cross section in the cold subhalos, present in our Galactic halo, with a mass hierarchy going all the way down to the damping mass limit of PeV dark matter, $m_{\rm min}$ of $\mathcal{O}$($10^{-18}$M$_\odot$). 

In this scenario, to obtain the observed PeV neutrino rate, it is sufficient to saturate the cross section at a value of $\mathcal{O}$(100) times the  {\it local} unitarity limit: $\left<\sigma_{\rm ann} v\right>_{\rm sat}\sim2.7\times10^{-23}$cm$^3$/s for $m_\chi=1$~PeV, at typical particle velocities of $\mathcal{O}$(1 km/s). A lower value of the {\it local} cross section would of course require a proportionally lower saturation velocity. The prediction in this model would be a signal with two main components: (i) a smooth dark matter contribution strongly peaked towards the Galactic Centre and (ii) and almost angle-independent contribution from dark matter subhalos where nearby large subhalos could appear as point sources. The relative contribution of these components across all angles would depend on the precise value of the velocity where the enhancement saturates. A lower saturation velocity would result in a stronger dominion of the subhalos. 

By looking at the all-sky distribution of events, we can see that a fraction of them ($\sim24\%$) are clustered around the Galactic Centre, although not at a strongly statistically significant level as pointed out before \cite{IceCube_2013}. Interestingly, 6 of the remaining 25 events coincide, within the angular errors, with the locations of five of the 26 MW satellites: Hercules, Sculptor, Sextans, Segue 1 and Ursa Major II. Although we have estimated that the probability of this (or more associated events) occurring randomly is $\sim35\%$, it would be worthy to test this possibility further once more events are collected.

Regarding the origin of PeV annihilating dark matter particles, we have also revised the possibility of being produced as thermal relics of the Big Bang. Although strong constraints on very heavy dark matter relics have been derived in the past, they have so far ignored the substantial reduction of the relic abundance due to the Sommerfeld mechanism after kinetic decoupling. In the extreme case where $T_{\rm kd}\sim T_f$, the standard relic abundance for a constant s-wave annihilation gets suppressed by a factor of  $0.2~{\rm ln}^{-1}(1/\sigma_{\rm vel}({\rm sat}))$, where $\sigma_{\rm vel}({\rm sat})$ is the 1D velocity dispersion of the dark matter particles when the Sommerfeld enhancement saturates. Even in this case however, the thermal relic abundance of PeV dark matter particles would overclose the Universe: $\Omega_\chi h^2\sim0.47$. A non-standard mechanism for dark matter production is therefore needed. For instance, if the Universe is reheated to a temperature $T_{\rm RH}\lesssim T_f$ by the decays of other unstable massive particles (e.g. moduli), then the relic abundance would be diluted by a factor of $(T_{\rm RH}/T_f)^3$.

In this paper we have considered only the case of annihilation {\it exclusively} to neutrinos $\chi\chi\rightarrow\nu{\bar \nu}$, i.e., at tree-level, there is only a coupling to neutrinos. This produces a monochromatic neutrino signal. In a broader scenario, annihilation into other channels would lead also to a continuum of lower energy cosmic ray neutrinos (from the decay of the primary annihilation byproducts). This could in principle explain the more numerous sub-PeV events reported by IceCube. The combination of a monochromatic line with a continuum might even explain the gap feature that exists in the observed spectra between $\sim0.4$~PeV and $\sim1$~PeV. For the case of DM decay, this has been shown explicitly (see e.g. Fig. 6 of \cite{Bai_13}). It would be interesting to study particle physics models with the required spectra and yield in the case of dark matter annihilation. We note that a possible neutrino continuum could be obtained by the electroweak radiative corrections to the $\chi\chi\rightarrow\nu{\bar \nu}$ process. This possibility  was studied in \cite{Serpico_07} where it was noted that the dominant $2\rightarrow3$ process is $\chi\chi\rightarrow\nu{\bar \nu}Z$. The authors estimated a branching ratio for this channel, $R=\sigma(\chi\chi\rightarrow\nu{\bar \nu}Z)/\sigma(\chi\chi\rightarrow\nu{\bar \nu})$, of $\mathcal{O}$(0.1) for $m_\chi\sim1$~PeV. Thus, a detailed analysis of this case might result in a non-negligible neutrino continuum.

However, once other byproducts of the annihilation are considered, it is important to keep in mind current astrophysical constraints. For instance, in the case considered above, there should be an associated diffuse gamma-ray signal from neutral pion decay produced by quark jets from $Z$ decays \cite{Serpico_07}. The spectra $E^2dN/dE$ of these gamma-rays however, would peak at energies probably too high to put any significant constraint with current experiments, $E_{\rm peak}\sim m_\chi/30\sim33$~TeV\footnote{We obtain this approximate value by noting that for annihilation into quark-antiquark pairs, or $W$ and $Z$ bosons, the continuous gamma-ray yield is approximated by the following formula: $dN/dE\sim(0.42/m_\chi){\rm exp}[-8x]/(x^{3/2}+1.4\times10^{-4})$, where $x=m_\chi/E$ (e.g. \cite{Bergstrom_01}).}. For instance, the stringent current gamma-ray constraints for dark matter annihilation from observations of the MW dwarf spheroidals (dSphs) stand at \cite{Fermi_13}:
\begin{equation}\label{const_fermi}
	 \left<\sigma_{	\rm ann} v\right>(\chi\chi\rightarrow b{\bar b}, m_\chi=10~{\rm TeV})<\mathcal{O}(10^{-23}{\rm cm}^3{\rm s}^{-1}).
\end{equation}
At higher dark matter masses, there are no constraints, but we note that the model we have considered here would even be consistent at the level of Eq.~(\ref{const_fermi}). This is because dSphs have typical velocities of $\mathcal{O}$(10 km/s), and thus, $\left<\sigma_{\rm ann} v\right>_{\rm dSphs}\sim2.7\times10^{-24}$cm$^3$/s in the example we explored in this paper.  

The associated gamma-rays from dark matter annihilation in extragalactic halos are attenuated by the opacity of the Universe caused by pair production with the Extragalactic Background Light and the Cosmic Microwave Background radiation. The resulting electron-positron pairs loose energy via Inverse Compton scattering with the photon backgrounds. The final result is a cascade of the original high energy
photons to lower GeV-TeV energies. This cascade is constrained by the extragalactic gamma-ray background observed by the {\it Fermi-LAT} instrument \cite{Abdo_2010}. In this way, it is possible to set an upper limit to the annihilation cross section in the channels that give rise to the original gamma-ray emission (e.g. \cite{Murase_2012}). Assuming a constant $\left<\sigma_{\rm ann} v\right>$ and a substructure boost more generous than the one we assumed here, this constraint stands at (see Fig. 15 of \cite{Murase_2012}):
\begin{equation}\label{const_cascade}
	 \left<\sigma_{	\rm ann} v\right>(\chi\chi\rightarrow [b{\bar b}~{\rm or}~W^{+}W^{-}~{\rm or}~\mu^+\mu^-], m_\chi=1~{\rm PeV})\lesssim3\times10^{-21}{\rm cm}^3{\rm s}^{-1}.
\end{equation}
This limit is satisfied by the maximum saturated cross section of the Sommerfeld-enhanced case studied here: $\left<\sigma_{\rm ann} v\right>_{\rm sat}\sim2.7\times10^{-23}$cm$^3$s$^{-1}$, which is
two orders of magnitude lower.

Another potential worry would be the energy injection in the early Universe due to dark matter annihilation. This could create distortions in the energy and power spectra of the CMB (e.g. \cite{Zavala_10,Slatyer_09}). The latter is the most constraining but still too weak at PeV masses to be of concern. The most recent analysis puts the following constraint \cite{Slatyer_13}:
\begin{equation}\label{constraint}
	p_{\rm ann}=\frac{f_{\rm eff}\left<\sigma_{\rm ann} v\right>}{m_\chi}<1.18\times10^{-27}{\rm cm}^3{\rm s}^{-1}{\rm GeV}^{-1},
\end{equation}
where $f_{\rm eff}$ is the efficiency factor to which the annihilation products get absorbed by the CMB plasma. For the case of a dominant channel of annihilation into neutrinos, most of the energy is lost and $f_{\rm eff}\ll1$. But even if one were to consider other annihilation channels and $f_{\rm eff}\sim1$, the constraint in Eq.~(\ref{constraint}) would be too weak for $m_\chi\sim1$~PeV.

\section*{Acknowledgments}

The Dark Cosmology Centre is funded by the DNRF. JZ is supported by the EU under a Marie Curie International Incoming Fellowship, contract PIIF-GA-2013-627723. I thank 
Niayesh Afshordi, Steen H. Hansen, Jens Hjorth, Jennifer Anne Adams, and the anonymous referee for useful comments and suggestions.

\appendix*

\section{Subhalo model}\label{sec_app}

If we assume that subhalos are a population of point sources in the sky, i.e., we neglect their spatial extent, then we can write the total $J-factor$ from substructures as \cite{Ando_09}:
\begin{equation}\label{Jfactor_subs}
	J_{\rm subs}(\Psi)=\frac{1}{\rho_\chi(R_{\odot})^2 L_{\rm MW}}\int_{\rm L_{min}}^{\rm L_{max}}dL\int_0^{\lambda_{\rm max}}L\frac{dn_{\rm sh}(\lambda,L)}{dL}d\lambda,
\end{equation}
where  $n_{\rm sh}$ is the radially dependent subhalo luminosity function:
\begin{equation}\label{lum_function}
	\frac{dn_{\rm sh}(r,L)}{dL}=\frac{dn_{\rm sh}(r,m_{\rm sub})}{dm_{\rm sub}}\frac{dm_{\rm sub}}{dL}=n_{\rm sh}(r)\frac{\alpha-1}{m_{\rm min}}\left(\frac{m_{\rm sub}}{m_{\rm min}}\right)^{-\alpha}\frac{dm_{\rm sub}}{dL},
\end{equation}
where $m_{\rm min}$ is the minimum subhalo mass corresponding to $L_{\rm min}$ in Eq.~(\ref{Jfactor_subs}).
We take analytical fits to this distribution from the Aq-A-1 MW halo simulation \cite{Springel_08}. The subhalo mass function has a slope of $\alpha=1.9$ and $n_{\rm sh}(r)$, the radial profile of the subhalo number density,
can be fitted by an Einasto profile:
\begin{eqnarray}\label{radial_dist}
	n_{\rm sh}(r)&=&\frac{f_{\rm sub}M_{\rm MW}}{2\pi r_{-2s }^3m_{\rm min}}\gamma\left(\frac{3}{\alpha_{es }},\frac{ 2c_{-2}^{\alpha_{es }} }{\alpha_{es }}\right)^{-1}
		\left(\frac{2}{\alpha_{es }}\right)^{3/\alpha_{es}}\nonumber\\
			&\times&{\rm exp}\left[\frac{-2}{\alpha_{es}}\left(\frac{r}{r_{-2s}}\right)^{\alpha_{es}}\right]\times\left(\frac{2-\alpha}{\alpha-1}\frac{1}{(m_{\rm max}/m_{\rm min})^{2-\alpha}-1}\right),
\end{eqnarray}
where $\alpha_{es}=0.678$, $r_{-2s}=0.81R_{200}$, $c_{-2}=r_{-2s}/R_{200}$, and $\gamma$ is the lower incomplete gamma function. Eq.~(\ref{radial_dist}) has been normalized so that the total mass in subhalos is a fraction $f_{\rm sub}$ of the virial mass of the simulated MW halo, $M_{\rm MW}\equiv M_{200}=1.41\times10^{12}$M$_\odot$\footnote{We note that $M_{200}$ is lower than the value given in Table 1 of \cite{Springel_08} because
we renormalized $\rho_\chi$ to the assumed local value: $\rho(R_\odot)=0.4$GeVcm$^{-3}$.}:
\begin{equation}
	M_{\rm subs}(R_{200})=4\pi\int_0^{R_{200}}r^2dr\int_{\rm m_{min}}^{\rm m_{max}}m_{\rm sub}\frac{dn_{\rm sh}(r,m_{\rm sub})}{dm_{\rm sub}}dm_{\rm sub}=f_{\rm sub}M_{\rm MW}.
\end{equation}
We take $m_{\rm max}=10^{10}$M$_\odot$, which is roughly the maximum subhalo mass in the Aq-A-1 simulation. Note that since we are assuming $\alpha=1.9$, the total subhalo mass in the limit $m_{\rm min}\rightarrow0$ converges. $M_{\rm subs}(R_{200})$ is actually almost converged at the resolution mass of the simulation, $m_{\rm res}\sim3\times10^4$M$_\odot$, with $f_{\rm sub}\sim0.13$. Unresolved substructures all the way down to $m_{\rm min}\rightarrow0$, only enhance $f_{\rm sub}$ by $\lesssim30\%$ \cite{Springel_08}. The mass contained in unresolved subhalos would be substantially larger if $\alpha$ was closer to
2. For this work we assume the value of $\alpha=1.9$ and take $f_{\rm sub}=0.13$.

We also note that subhalos are distributed radially in a way which is considerably shallower than the smooth distribution. This is due to tidal stripping that disrupts subhalos in the central dense regions of the host halo. This disruption seems to occur in such a way that the radial dependence of the subhalo distribution is independent of mass. This is observed in numerical simulations (e.g. Fig. 11 of \cite{Springel_08}). We assume that this radial dependence holds down for lower unresolved masses. This is expected since although halos with smaller masses collapse earlier, and thus are denser and more resilient to tidal stripping that more massive halos, the epochs of collapse are not substantially different due to the flattening of the power spectrum at smaller scales.

Assuming that each subhalo can be represented by a spherical distribution of dark matter with a radial NFW profile \cite{NFW_97}\footnote{Subhalos are better fitted by Einasto profiles, but the simplicity of the NFW profile makes it convenient for our calculations. It is for example not clear what is the dependence of the Einasto parameters with subhalo mass. Using an Einasto profile instead of a NFW actually increases the net annihilation rate in
a halo by $\sim50\%$ (see e.g. \cite{Zavala_10}).}, their individual luminosities are given by:
\begin{equation}\label{l_sub}
	L=\int_0^{r_{200}}\rho_{\rm NFW}^2(r)dV=1.23\frac{V_{\rm max}^4}{G^2r_{\rm max}}\left(1-\frac{1}{(1+c)^3}\right),
\end{equation}
where $c=r_{200}/r_s$ is the concentration of the subhalo ($r_s$ being the scale radius in the NFW profile), and $r_{\rm max}$ is the radius where the circular velocity reaches its maximum $V_{\rm max}$. The term in parentheses in Eq.~(\ref{l_sub}) comes from truncating the integral to the virial radius of the subhalo $r_{200}$. In principle, a more appropriate truncation radius should be the tidal radius, which would depend on the gravitational potential of the host halo. However, due to the $\rho^2$ dependence of the luminosity, most of the annihilation occurs in the very central regions,  $L(r<r_{1/2}=0.25r_s)=L/2$. Since $r_{\rm max}=2.163r_s$ for the NFW profile, then $r_{1/2}\sim0.1r_{\rm max}$ which is significantly smaller than typical tidal radii, except in the cases of extreme disruption. We therefore use Eq.~(\ref{l_sub}) noting that a proper truncation would decrease the luminosities by a factor $\ll0.5$.

The scaling properties of the subhalos are tightly correlated to the subhalo mass. We take the mean correlations computed from the distribution of resolved subhalos in the Aq-A-1 simulation, without considering the spread of these distributions. We have:
\begin{eqnarray}\label{scaling}
	V_{\rm max}&=&10~{\rm km/s}\left(\frac{m_{\rm sub}}{3.37\times10^7{\rm M}_\odot}\right)^{1/3.49}\nonumber\\
	r_{\rm max}&=&5.87\times10^{-3}\left(\frac{V_{\rm max}}{{\rm H_0}}\right)\left(\frac{m_{\rm sub}}{10^8{\rm M}_\odot}\right)^{0.09},
\end{eqnarray}
where $H_0=100$~km~s$^{-1}$~Mpc$^{-1}h$. The subhalo concentration is then simply given by solving the transcendental equation \cite{Springel_08}: 
\begin{equation}\label{delta_c}
	\delta_c=\frac{200}{3}\frac{c^3}{{\rm ln}(1+c) - c/(1+c)}=7.213~\delta_V=7.213\frac{\bar{\rho}(r_{\rm max})}{\rho_{\rm crit}}=14.426\left(\frac{V_{\rm max}}{H_0 r_{\rm max}}\right)^2,
\end{equation}
where $\delta_V$ is the mean overdensity within $r_{\rm max}$ relative to the critical density. 

These scaling relations result in a concentration-mass relation that is well fitted by a power law. However, it is not appropriate to extrapolate this power law down to unresolved masses since, as discussed above, the flattening of the CDM power spectrum at lower masses implies a flattening of the concentration-mass relation, which considerably reduces the unresolved subhalo contribution \cite{Zavala_13b,MASC_13}. To account for this effect, we use the fitting function recently proposed by \cite{MASC_13}:
\begin{equation}\label{c_fit}
	c_{\rm fit}=\sum_{i=0}^{i=5}c_i\times\left[{\rm ln}\left(\frac{m}{{\rm M}_\odot h^{-1}}\right)\right]^i
\end{equation}
where $c_i=(37.5153, -1.5093, 1.636\times10^{-2}, 3.66\times10^{-4},-2.89237\times10^{-5}, 5.32\times10^{-7})$. Since this formula is strictly valid only for field main halos, we re-normalize it to match the concentration-mass relation implied in Eqs.~(\ref{scaling}-\ref{delta_c}) above for subhalos. It is known that subhalo concentrations are biased towards higher values roughly by the same factor across different masses (e.g., see Fig. 26 of \cite{Springel_08}).

Finally we note that although in principle the full hierarchy of sub-substructures should be considered to estimate the total subhalo contribution, the first level of the hierarchy is the dominant one since further levels, most abundant in the outskirts of subhalos, would be removed rapidly by tidal stripping with the host in the first orbital interactions. For subhalos that are still at first infall and near the virial radius of the host, sub-substructures might survive in significant numbers to contribute to the annihilation emission. This would enhance the number of neutrino events estimated here.

With the whole set of Eqs.~(\ref{lum_function}-\ref{c_fit}) we can therefore estimate the contribution from substructure to the dark matter annihilation rate.

\subsection{Sommerfeld enhancement for subhalos}

We use a similar approach to the one we used for the smooth dark matter component (see Section~\ref{sec_SE_main}), but make a further simplification and take the {\it average} 1D velocity dispersion of each substructure, $\bar{\sigma}_{\rm vel}\sim V_{\rm max}/\sqrt{3}$, as a measure of the enhancement, i.e., individual subhalo luminosities (Eq.~\ref{l_sub}) get enhanced by:
\begin{equation}
	L(m_{\rm sub})\rightarrow \left(\frac{1}{  \sqrt{\pi}\bar{\sigma}_{\rm vel} }\right)L(m_{\rm sub})
\end{equation}

\bibliography{lit}

\begin{thebibliography}{50}
\expandafter\ifx\csname natexlab\endcsname\relax\def\natexlab#1{#1}\fi
\expandafter\ifx\csname bibnamefont\endcsname\relax
  \def\bibnamefont#1{#1}\fi
\expandafter\ifx\csname bibfnamefont\endcsname\relax
  \def\bibfnamefont#1{#1}\fi
\expandafter\ifx\csname citenamefont\endcsname\relax
  \def\citenamefont#1{#1}\fi
\expandafter\ifx\csname url\endcsname\relax
  \def\url#1{\texttt{#1}}\fi
\expandafter\ifx\csname urlprefix\endcsname\relax\def\urlprefix{URL }\fi
\providecommand{\bibinfo}[2]{#2}
\providecommand{\eprint}[2][]{\url{#2}}

\bibitem[{\citenamefont{{IceCube Collaboration}}(2013)}]{IceCube_2013}
\bibinfo{author}{\bibnamefont{{IceCube Collaboration}}},
  \bibinfo{journal}{Science} \textbf{\bibinfo{volume}{342}}
  (\bibinfo{year}{2013}), \eprint{1311.5238}.

\bibitem[{\citenamefont{{Aartsen} et~al.}(2013)\citenamefont{{Aartsen},
  {Abbasi}, {Abdou}, {Ackermann}, {Adams}, {Aguilar}, {Ahlers}, {Altmann},
  {Auffenberg}, {Bai} et~al.}}]{IceCube_2013_2}
\bibinfo{author}{\bibfnamefont{M.~G.} \bibnamefont{{Aartsen}}},
  \bibinfo{author}{\bibfnamefont{R.}~\bibnamefont{{Abbasi}}},
  \bibinfo{author}{\bibfnamefont{Y.}~\bibnamefont{{Abdou}}},
  \bibinfo{author}{\bibfnamefont{M.}~\bibnamefont{{Ackermann}}},
  \bibinfo{author}{\bibfnamefont{J.}~\bibnamefont{{Adams}}},
  \bibinfo{author}{\bibfnamefont{J.~A.} \bibnamefont{{Aguilar}}},
  \bibinfo{author}{\bibfnamefont{M.}~\bibnamefont{{Ahlers}}},
  \bibinfo{author}{\bibfnamefont{D.}~\bibnamefont{{Altmann}}},
  \bibinfo{author}{\bibfnamefont{J.}~\bibnamefont{{Auffenberg}}},
  \bibinfo{author}{\bibfnamefont{X.}~\bibnamefont{{Bai}}},
  \bibnamefont{et~al.}, \bibinfo{journal}{Physical Review Letters}
  \textbf{\bibinfo{volume}{111}}, \bibinfo{eid}{021103} (\bibinfo{year}{2013}).

\bibitem[{\citenamefont{{Aartsen} et~al.}(2014)\citenamefont{{Aartsen},
  {Ackermann}, {Adams}, {Aguilar}, {Ahlers}, {Ahrens}, {Altmann}, {Anderson},
  {Arguelles}, {Arlen} et~al.}}]{IceCube_2014}
\bibinfo{author}{\bibfnamefont{M.~G.} \bibnamefont{{Aartsen}}},
  \bibinfo{author}{\bibfnamefont{M.}~\bibnamefont{{Ackermann}}},
  \bibinfo{author}{\bibfnamefont{J.}~\bibnamefont{{Adams}}},
  \bibinfo{author}{\bibfnamefont{J.~A.} \bibnamefont{{Aguilar}}},
  \bibinfo{author}{\bibfnamefont{M.}~\bibnamefont{{Ahlers}}},
  \bibinfo{author}{\bibfnamefont{M.}~\bibnamefont{{Ahrens}}},
  \bibinfo{author}{\bibfnamefont{D.}~\bibnamefont{{Altmann}}},
  \bibinfo{author}{\bibfnamefont{T.}~\bibnamefont{{Anderson}}},
  \bibinfo{author}{\bibfnamefont{C.}~\bibnamefont{{Arguelles}}},
  \bibinfo{author}{\bibfnamefont{T.~C.} \bibnamefont{{Arlen}}},
  \bibnamefont{et~al.}, \bibinfo{journal}{ArXiv e-prints}
  (\bibinfo{year}{2014}), \eprint{1405.5303}.

\bibitem[{\citenamefont{{Anchordoqui} et~al.}(2013)\citenamefont{{Anchordoqui},
  {Barger}, {Cholis}, {Goldberg}, {Hooper}, {Kusenko}, {Learned}, {Marfatia},
  {Pakvasa}, {Paul} et~al.}}]{Cosmic_PeV}
\bibinfo{author}{\bibfnamefont{L.~A.} \bibnamefont{{Anchordoqui}}},
  \bibinfo{author}{\bibfnamefont{V.}~\bibnamefont{{Barger}}},
  \bibinfo{author}{\bibfnamefont{I.}~\bibnamefont{{Cholis}}},
  \bibinfo{author}{\bibfnamefont{H.}~\bibnamefont{{Goldberg}}},
  \bibinfo{author}{\bibfnamefont{D.}~\bibnamefont{{Hooper}}},
  \bibinfo{author}{\bibfnamefont{A.}~\bibnamefont{{Kusenko}}},
  \bibinfo{author}{\bibfnamefont{J.~G.} \bibnamefont{{Learned}}},
  \bibinfo{author}{\bibfnamefont{D.}~\bibnamefont{{Marfatia}}},
  \bibinfo{author}{\bibfnamefont{S.}~\bibnamefont{{Pakvasa}}},
  \bibinfo{author}{\bibfnamefont{T.~C.} \bibnamefont{{Paul}}},
  \bibnamefont{et~al.}, \bibinfo{journal}{ArXiv e-prints}
  (\bibinfo{year}{2013}), \eprint{1312.6587}.

\bibitem[{\citenamefont{{Feldstein} et~al.}(2013)\citenamefont{{Feldstein},
  {Kusenko}, {Matsumoto}, and {Yanagida}}}]{Feldstein_13}
\bibinfo{author}{\bibfnamefont{B.}~\bibnamefont{{Feldstein}}},
  \bibinfo{author}{\bibfnamefont{A.}~\bibnamefont{{Kusenko}}},
  \bibinfo{author}{\bibfnamefont{S.}~\bibnamefont{{Matsumoto}}},
  \bibnamefont{and} \bibinfo{author}{\bibfnamefont{T.~T.}
  \bibnamefont{{Yanagida}}}, \bibinfo{journal}{\prd}
  \textbf{\bibinfo{volume}{88}}, \bibinfo{eid}{015004} (\bibinfo{year}{2013}),
  \eprint{1303.7320}.

\bibitem[{\citenamefont{{Esmaili} and {Dario Serpico}}(2013)}]{Esmaili_13}
\bibinfo{author}{\bibfnamefont{A.}~\bibnamefont{{Esmaili}}} \bibnamefont{and}
  \bibinfo{author}{\bibfnamefont{P.}~\bibnamefont{{Dario Serpico}}},
  \bibinfo{journal}{Journal of Cosmology and Astroparticle Physics}
  \textbf{\bibinfo{volume}{11}}, \bibinfo{eid}{054} (\bibinfo{year}{2013}),
  \eprint{1308.1105}.

\bibitem[{\citenamefont{{Bai} et~al.}(2013)\citenamefont{{Bai}, {Lu}, and
  {Salvado}}}]{Bai_13}
\bibinfo{author}{\bibfnamefont{Y.}~\bibnamefont{{Bai}}},
  \bibinfo{author}{\bibfnamefont{R.}~\bibnamefont{{Lu}}}, \bibnamefont{and}
  \bibinfo{author}{\bibfnamefont{J.}~\bibnamefont{{Salvado}}},
  \bibinfo{journal}{ArXiv e-prints}  (\bibinfo{year}{2013}),
  \eprint{1311.5864}.

\bibitem[{\citenamefont{{Bhattacharya}
  et~al.}(2014)\citenamefont{{Bhattacharya}, {Hall Reno}, and
  {Sarcevic}}}]{Bat_14}
\bibinfo{author}{\bibfnamefont{A.}~\bibnamefont{{Bhattacharya}}},
  \bibinfo{author}{\bibfnamefont{M.}~\bibnamefont{{Hall Reno}}},
  \bibnamefont{and}
  \bibinfo{author}{\bibfnamefont{I.}~\bibnamefont{{Sarcevic}}},
  \bibinfo{journal}{ArXiv e-prints}  (\bibinfo{year}{2014}),
  \eprint{1403.1862}.

\bibitem[{\citenamefont{{Gandhi} et~al.}(1998)\citenamefont{{Gandhi}, {Quigg},
  {Reno}, and {Sarcevic}}}]{Gandhi_98}
\bibinfo{author}{\bibfnamefont{R.}~\bibnamefont{{Gandhi}}},
  \bibinfo{author}{\bibfnamefont{C.}~\bibnamefont{{Quigg}}},
  \bibinfo{author}{\bibfnamefont{M.~H.} \bibnamefont{{Reno}}},
  \bibnamefont{and}
  \bibinfo{author}{\bibfnamefont{I.}~\bibnamefont{{Sarcevic}}},
  \bibinfo{journal}{\prd} \textbf{\bibinfo{volume}{58}}, \bibinfo{eid}{093009}
  (\bibinfo{year}{1998}), \eprint{hep-ph/9807264}.

\bibitem[{\citenamefont{{Bovy} and {Tremaine}}(2012{\natexlab{a}})}]{Bovy_12}
\bibinfo{author}{\bibfnamefont{J.}~\bibnamefont{{Bovy}}} \bibnamefont{and}
  \bibinfo{author}{\bibfnamefont{S.}~\bibnamefont{{Tremaine}}},
  \bibinfo{journal}{\apj} \textbf{\bibinfo{volume}{756}}, \bibinfo{eid}{89}
  (\bibinfo{year}{2012}{\natexlab{a}}), \eprint{1205.4033}.

\bibitem[{\citenamefont{{Hisano} et~al.}(2004)\citenamefont{{Hisano},
  {Matsumoto}, and {Nojiri}}}]{Hisano_04}
\bibinfo{author}{\bibfnamefont{J.}~\bibnamefont{{Hisano}}},
  \bibinfo{author}{\bibfnamefont{S.}~\bibnamefont{{Matsumoto}}},
  \bibnamefont{and} \bibinfo{author}{\bibfnamefont{M.~M.}
  \bibnamefont{{Nojiri}}}, \bibinfo{journal}{Physical Review Letters}
  \textbf{\bibinfo{volume}{92}}, \bibinfo{pages}{031303}
  (\bibinfo{year}{2004}), \eprint{arXiv:hep-ph/0307216}.

\bibitem[{\citenamefont{{Arkani-Hamed}
  et~al.}(2009)\citenamefont{{Arkani-Hamed}, {Finkbeiner}, {Slatyer}, and
  {Weiner}}}]{Arkani_09}
\bibinfo{author}{\bibfnamefont{N.}~\bibnamefont{{Arkani-Hamed}}},
  \bibinfo{author}{\bibfnamefont{D.~P.} \bibnamefont{{Finkbeiner}}},
  \bibinfo{author}{\bibfnamefont{T.~R.} \bibnamefont{{Slatyer}}},
  \bibnamefont{and} \bibinfo{author}{\bibfnamefont{N.}~\bibnamefont{{Weiner}}},
  \bibinfo{journal}{\prd} \textbf{\bibinfo{volume}{79}},
  \bibinfo{pages}{015014} (\bibinfo{year}{2009}), \eprint{0810.0713}.

\bibitem[{\citenamefont{{Lattanzi} and {Silk}}(2009)}]{Lattanzi_09}
\bibinfo{author}{\bibfnamefont{M.}~\bibnamefont{{Lattanzi}}} \bibnamefont{and}
  \bibinfo{author}{\bibfnamefont{J.}~\bibnamefont{{Silk}}},
  \bibinfo{journal}{\prd} \textbf{\bibinfo{volume}{79}},
  \bibinfo{pages}{083523} (\bibinfo{year}{2009}), \eprint{0812.0360}.

\bibitem[{\citenamefont{{Y{\"u}ksel} et~al.}(2007)\citenamefont{{Y{\"u}ksel},
  {Horiuchi}, {Beacom}, and {Ando}}}]{Yuksel_07}
\bibinfo{author}{\bibfnamefont{H.}~\bibnamefont{{Y{\"u}ksel}}},
  \bibinfo{author}{\bibfnamefont{S.}~\bibnamefont{{Horiuchi}}},
  \bibinfo{author}{\bibfnamefont{J.~F.} \bibnamefont{{Beacom}}},
  \bibnamefont{and} \bibinfo{author}{\bibfnamefont{S.}~\bibnamefont{{Ando}}},
  \bibinfo{journal}{\prd} \textbf{\bibinfo{volume}{76}}, \bibinfo{eid}{123506}
  (\bibinfo{year}{2007}), \eprint{0707.0196}.

\bibitem[{\citenamefont{{Springel}
  et~al.}(2008{\natexlab{a}})\citenamefont{{Springel}, {Wang}, {Vogelsberger},
  {Ludlow}, {Jenkins}, {Helmi}, {Navarro}, {Frenk}, and {White}}}]{Springel_08}
\bibinfo{author}{\bibfnamefont{V.}~\bibnamefont{{Springel}}},
  \bibinfo{author}{\bibfnamefont{J.}~\bibnamefont{{Wang}}},
  \bibinfo{author}{\bibfnamefont{M.}~\bibnamefont{{Vogelsberger}}},
  \bibinfo{author}{\bibfnamefont{A.}~\bibnamefont{{Ludlow}}},
  \bibinfo{author}{\bibfnamefont{A.}~\bibnamefont{{Jenkins}}},
  \bibinfo{author}{\bibfnamefont{A.}~\bibnamefont{{Helmi}}},
  \bibinfo{author}{\bibfnamefont{J.~F.} \bibnamefont{{Navarro}}},
  \bibinfo{author}{\bibfnamefont{C.~S.} \bibnamefont{{Frenk}}},
  \bibnamefont{and} \bibinfo{author}{\bibfnamefont{S.~D.~M.}
  \bibnamefont{{White}}}, \bibinfo{journal}{\mnras}
  \textbf{\bibinfo{volume}{391}}, \bibinfo{pages}{1685}
  (\bibinfo{year}{2008}{\natexlab{a}}), \eprint{0809.0898}.

\bibitem[{\citenamefont{{Navarro} et~al.}(2010)\citenamefont{{Navarro},
  {Ludlow}, {Springel}, {Wang}, {Vogelsberger}, {White}, {Jenkins}, {Frenk},
  and {Helmi}}}]{Navarro_10}
\bibinfo{author}{\bibfnamefont{J.~F.} \bibnamefont{{Navarro}}},
  \bibinfo{author}{\bibfnamefont{A.}~\bibnamefont{{Ludlow}}},
  \bibinfo{author}{\bibfnamefont{V.}~\bibnamefont{{Springel}}},
  \bibinfo{author}{\bibfnamefont{J.}~\bibnamefont{{Wang}}},
  \bibinfo{author}{\bibfnamefont{M.}~\bibnamefont{{Vogelsberger}}},
  \bibinfo{author}{\bibfnamefont{S.~D.~M.} \bibnamefont{{White}}},
  \bibinfo{author}{\bibfnamefont{A.}~\bibnamefont{{Jenkins}}},
  \bibinfo{author}{\bibfnamefont{C.~S.} \bibnamefont{{Frenk}}},
  \bibnamefont{and} \bibinfo{author}{\bibfnamefont{A.}~\bibnamefont{{Helmi}}},
  \bibinfo{journal}{\mnras} \textbf{\bibinfo{volume}{402}}, \bibinfo{pages}{21}
  (\bibinfo{year}{2010}), \eprint{0810.1522}.

\bibitem[{\citenamefont{{Bovy} and {Tremaine}}(2012{\natexlab{b}})}]{Bovy_2012}
\bibinfo{author}{\bibfnamefont{J.}~\bibnamefont{{Bovy}}} \bibnamefont{and}
  \bibinfo{author}{\bibfnamefont{S.}~\bibnamefont{{Tremaine}}},
  \bibinfo{journal}{\apj} \textbf{\bibinfo{volume}{756}}, \bibinfo{eid}{89}
  (\bibinfo{year}{2012}{\natexlab{b}}), \eprint{1205.4033}.

\bibitem[{\citenamefont{{Boylan-Kolchin}
  et~al.}(2012)\citenamefont{{Boylan-Kolchin}, {Bullock}, and
  {Kaplinghat}}}]{BK_2012}
\bibinfo{author}{\bibfnamefont{M.}~\bibnamefont{{Boylan-Kolchin}}},
  \bibinfo{author}{\bibfnamefont{J.~S.} \bibnamefont{{Bullock}}},
  \bibnamefont{and}
  \bibinfo{author}{\bibfnamefont{M.}~\bibnamefont{{Kaplinghat}}},
  \bibinfo{journal}{\mnras} \textbf{\bibinfo{volume}{422}},
  \bibinfo{pages}{1203} (\bibinfo{year}{2012}), \eprint{1111.2048}.

\bibitem[{\citenamefont{{Bovy}}(2009)}]{Bovy_09}
\bibinfo{author}{\bibfnamefont{J.}~\bibnamefont{{Bovy}}},
  \bibinfo{journal}{\prd} \textbf{\bibinfo{volume}{79}}, \bibinfo{eid}{083539}
  (\bibinfo{year}{2009}), \eprint{0903.0413}.

\bibitem[{\citenamefont{{Zavala} and
  {Afshordi}}(2013{\natexlab{a}})}]{Zavala_13a}
\bibinfo{author}{\bibfnamefont{J.}~\bibnamefont{{Zavala}}} \bibnamefont{and}
  \bibinfo{author}{\bibfnamefont{N.}~\bibnamefont{{Afshordi}}},
  \bibinfo{journal}{ArXiv e-prints}  (\bibinfo{year}{2013}{\natexlab{a}}),
  \eprint{1308.1098}.

\bibitem[{\citenamefont{{Ferrer} and {Hunter}}(2013)}]{Ferrer_13}
\bibinfo{author}{\bibfnamefont{F.}~\bibnamefont{{Ferrer}}} \bibnamefont{and}
  \bibinfo{author}{\bibfnamefont{D.~R.} \bibnamefont{{Hunter}}},
  \bibinfo{journal}{Journal of Cosmology and Astroparticle Physics}
  \textbf{\bibinfo{volume}{9}}, \bibinfo{eid}{005} (\bibinfo{year}{2013}),
  \eprint{1306.6586}.

\bibitem[{\citenamefont{{Zavala} and
  {Afshordi}}(2013{\natexlab{b}})}]{Zavala_13b}
\bibinfo{author}{\bibfnamefont{J.}~\bibnamefont{{Zavala}}} \bibnamefont{and}
  \bibinfo{author}{\bibfnamefont{N.}~\bibnamefont{{Afshordi}}},
  \bibinfo{journal}{ArXiv e-prints}  (\bibinfo{year}{2013}{\natexlab{b}}),
  \eprint{1311.3296}.

\bibitem[{\citenamefont{{Davis} and {Peebles}}(1977)}]{Davis_77}
\bibinfo{author}{\bibfnamefont{M.}~\bibnamefont{{Davis}}} \bibnamefont{and}
  \bibinfo{author}{\bibfnamefont{P.~J.~E.} \bibnamefont{{Peebles}}},
  \bibinfo{journal}{\apjs} \textbf{\bibinfo{volume}{34}}, \bibinfo{pages}{425}
  (\bibinfo{year}{1977}).

\bibitem[{\citenamefont{{Afshordi} et~al.}(2010)\citenamefont{{Afshordi},
  {Mohayaee}, and {Bertschinger}}}]{Afshordi_10}
\bibinfo{author}{\bibfnamefont{N.}~\bibnamefont{{Afshordi}}},
  \bibinfo{author}{\bibfnamefont{R.}~\bibnamefont{{Mohayaee}}},
  \bibnamefont{and}
  \bibinfo{author}{\bibfnamefont{E.}~\bibnamefont{{Bertschinger}}},
  \bibinfo{journal}{\prd} \textbf{\bibinfo{volume}{81}}, \bibinfo{eid}{101301}
  (\bibinfo{year}{2010}), \eprint{0911.0414}.

\bibitem[{\citenamefont{{Springel}
  et~al.}(2008{\natexlab{b}})\citenamefont{{Springel}, {White}, {Frenk},
  {Navarro}, {Jenkins}, {Vogelsberger}, {Wang}, {Ludlow}, and
  {Helmi}}}]{Springel_08_2}
\bibinfo{author}{\bibfnamefont{V.}~\bibnamefont{{Springel}}},
  \bibinfo{author}{\bibfnamefont{S.~D.~M.} \bibnamefont{{White}}},
  \bibinfo{author}{\bibfnamefont{C.~S.} \bibnamefont{{Frenk}}},
  \bibinfo{author}{\bibfnamefont{J.~F.} \bibnamefont{{Navarro}}},
  \bibinfo{author}{\bibfnamefont{A.}~\bibnamefont{{Jenkins}}},
  \bibinfo{author}{\bibfnamefont{M.}~\bibnamefont{{Vogelsberger}}},
  \bibinfo{author}{\bibfnamefont{J.}~\bibnamefont{{Wang}}},
  \bibinfo{author}{\bibfnamefont{A.}~\bibnamefont{{Ludlow}}}, \bibnamefont{and}
  \bibinfo{author}{\bibfnamefont{A.}~\bibnamefont{{Helmi}}},
  \bibinfo{journal}{nat} \textbf{\bibinfo{volume}{456}}, \bibinfo{pages}{73}
  (\bibinfo{year}{2008}{\natexlab{b}}), \eprint{0809.0894}.

\bibitem[{\citenamefont{{Fornasa} et~al.}(2013)\citenamefont{{Fornasa},
  {Zavala}, {S{\'a}nchez-Conde}, {Siegal-Gaskins}, {Delahaye}, {Prada},
  {Vogelsberger}, {Zandanel}, and {Frenk}}}]{Fornasa_13}
\bibinfo{author}{\bibfnamefont{M.}~\bibnamefont{{Fornasa}}},
  \bibinfo{author}{\bibfnamefont{J.}~\bibnamefont{{Zavala}}},
  \bibinfo{author}{\bibfnamefont{M.~A.} \bibnamefont{{S{\'a}nchez-Conde}}},
  \bibinfo{author}{\bibfnamefont{J.~M.} \bibnamefont{{Siegal-Gaskins}}},
  \bibinfo{author}{\bibfnamefont{T.}~\bibnamefont{{Delahaye}}},
  \bibinfo{author}{\bibfnamefont{F.}~\bibnamefont{{Prada}}},
  \bibinfo{author}{\bibfnamefont{M.}~\bibnamefont{{Vogelsberger}}},
  \bibinfo{author}{\bibfnamefont{F.}~\bibnamefont{{Zandanel}}},
  \bibnamefont{and} \bibinfo{author}{\bibfnamefont{C.~S.}
  \bibnamefont{{Frenk}}}, \bibinfo{journal}{\mnras}
  \textbf{\bibinfo{volume}{429}}, \bibinfo{pages}{1529} (\bibinfo{year}{2013}),
  \eprint{1207.0502}.

\bibitem[{\citenamefont{{Gehrels}}(1986)}]{Gehrels_1986}
\bibinfo{author}{\bibfnamefont{N.}~\bibnamefont{{Gehrels}}},
  \bibinfo{journal}{\apj} \textbf{\bibinfo{volume}{303}}, \bibinfo{pages}{336}
  (\bibinfo{year}{1986}).

\bibitem[{\citenamefont{{Kuhlen} et~al.}(2009)\citenamefont{{Kuhlen}, {Madau},
  and {Silk}}}]{Kuhlen_09}
\bibinfo{author}{\bibfnamefont{M.}~\bibnamefont{{Kuhlen}}},
  \bibinfo{author}{\bibfnamefont{P.}~\bibnamefont{{Madau}}}, \bibnamefont{and}
  \bibinfo{author}{\bibfnamefont{J.}~\bibnamefont{{Silk}}},
  \bibinfo{journal}{Science} \textbf{\bibinfo{volume}{325}},
  \bibinfo{pages}{970} (\bibinfo{year}{2009}), \eprint{0907.0005}.

\bibitem[{\citenamefont{{The Fermi-LAT Collaboration}
  et~al.}(2013)\citenamefont{{The Fermi-LAT Collaboration}, {:}, {Ackermann},
  {Albert}, {Anderson}, {Baldini}, {Ballet}, {Barbiellini}, {Bastieri},
  {Bechtol} et~al.}}]{Fermi_13}
\bibinfo{author}{\bibnamefont{{The Fermi-LAT Collaboration}}},
  \bibinfo{author}{\bibnamefont{{:}}},
  \bibinfo{author}{\bibfnamefont{M.}~\bibnamefont{{Ackermann}}},
  \bibinfo{author}{\bibfnamefont{A.}~\bibnamefont{{Albert}}},
  \bibinfo{author}{\bibfnamefont{B.}~\bibnamefont{{Anderson}}},
  \bibinfo{author}{\bibfnamefont{L.}~\bibnamefont{{Baldini}}},
  \bibinfo{author}{\bibfnamefont{J.}~\bibnamefont{{Ballet}}},
  \bibinfo{author}{\bibfnamefont{G.}~\bibnamefont{{Barbiellini}}},
  \bibinfo{author}{\bibfnamefont{D.}~\bibnamefont{{Bastieri}}},
  \bibinfo{author}{\bibfnamefont{K.}~\bibnamefont{{Bechtol}}},
  \bibnamefont{et~al.}, \bibinfo{journal}{ArXiv e-prints}
  (\bibinfo{year}{2013}), \eprint{1310.0828}.

\bibitem[{\citenamefont{{Walker} et~al.}(2009)\citenamefont{{Walker}, {Mateo},
  {Olszewski}, {Pe{\~n}arrubia}, {Wyn Evans}, and {Gilmore}}}]{Walker_2009}
\bibinfo{author}{\bibfnamefont{M.~G.} \bibnamefont{{Walker}}},
  \bibinfo{author}{\bibfnamefont{M.}~\bibnamefont{{Mateo}}},
  \bibinfo{author}{\bibfnamefont{E.~W.} \bibnamefont{{Olszewski}}},
  \bibinfo{author}{\bibfnamefont{J.}~\bibnamefont{{Pe{\~n}arrubia}}},
  \bibinfo{author}{\bibfnamefont{N.}~\bibnamefont{{Wyn Evans}}},
  \bibnamefont{and}
  \bibinfo{author}{\bibfnamefont{G.}~\bibnamefont{{Gilmore}}},
  \bibinfo{journal}{\apj} \textbf{\bibinfo{volume}{704}}, \bibinfo{pages}{1274}
  (\bibinfo{year}{2009}), \eprint{0906.0341}.

\bibitem[{\citenamefont{{Navarro} et~al.}(1997)\citenamefont{{Navarro},
  {Frenk}, and {White}}}]{NFW_97}
\bibinfo{author}{\bibfnamefont{J.~F.} \bibnamefont{{Navarro}}},
  \bibinfo{author}{\bibfnamefont{C.~S.} \bibnamefont{{Frenk}}},
  \bibnamefont{and} \bibinfo{author}{\bibfnamefont{S.~D.~M.}
  \bibnamefont{{White}}}, \bibinfo{journal}{\apj}
  \textbf{\bibinfo{volume}{490}}, \bibinfo{pages}{493} (\bibinfo{year}{1997}),
  \eprint{astro-ph/9611107}.

\bibitem[{\citenamefont{{Bovy} and {Rix}}(2013)}]{Bovy_2013}
\bibinfo{author}{\bibfnamefont{J.}~\bibnamefont{{Bovy}}} \bibnamefont{and}
  \bibinfo{author}{\bibfnamefont{H.-W.} \bibnamefont{{Rix}}},
  \bibinfo{journal}{\apj} \textbf{\bibinfo{volume}{779}}, \bibinfo{eid}{115}
  (\bibinfo{year}{2013}), \eprint{1309.0809}.

\bibitem[{\citenamefont{{Nesti} and {Salucci}}(2013)}]{Salucci_2013}
\bibinfo{author}{\bibfnamefont{F.}~\bibnamefont{{Nesti}}} \bibnamefont{and}
  \bibinfo{author}{\bibfnamefont{P.}~\bibnamefont{{Salucci}}},
  \bibinfo{journal}{Journal of Cosmology and Astroparticle Physics}
  \textbf{\bibinfo{volume}{7}}, \bibinfo{eid}{016} (\bibinfo{year}{2013}),
  \eprint{1304.5127}.

\bibitem[{\citenamefont{{S{\'a}nchez-Conde}
  et~al.}(2011)\citenamefont{{S{\'a}nchez-Conde}, {Cannoni}, {Zandanel},
  {G{\'o}mez}, and {Prada}}}]{MASC_2011}
\bibinfo{author}{\bibfnamefont{M.~A.} \bibnamefont{{S{\'a}nchez-Conde}}},
  \bibinfo{author}{\bibfnamefont{M.}~\bibnamefont{{Cannoni}}},
  \bibinfo{author}{\bibfnamefont{F.}~\bibnamefont{{Zandanel}}},
  \bibinfo{author}{\bibfnamefont{M.~E.} \bibnamefont{{G{\'o}mez}}},
  \bibnamefont{and} \bibinfo{author}{\bibfnamefont{F.}~\bibnamefont{{Prada}}},
  \bibinfo{journal}{Journal of Cosmology and Astroparticle Physics}
  \textbf{\bibinfo{volume}{12}}, \bibinfo{eid}{011} (\bibinfo{year}{2011}),
  \eprint{1104.3530}.

\bibitem[{\citenamefont{{Griest} and {Kamionkowski}}(1990)}]{Griest_90}
\bibinfo{author}{\bibfnamefont{K.}~\bibnamefont{{Griest}}} \bibnamefont{and}
  \bibinfo{author}{\bibfnamefont{M.}~\bibnamefont{{Kamionkowski}}},
  \bibinfo{journal}{Physical Review Letters} \textbf{\bibinfo{volume}{64}},
  \bibinfo{pages}{615} (\bibinfo{year}{1990}).

\bibitem[{\citenamefont{{Zavala}
  et~al.}(2010{\natexlab{a}})\citenamefont{{Zavala}, {Vogelsberger}, and
  {White}}}]{Zavala_10_2}
\bibinfo{author}{\bibfnamefont{J.}~\bibnamefont{{Zavala}}},
  \bibinfo{author}{\bibfnamefont{M.}~\bibnamefont{{Vogelsberger}}},
  \bibnamefont{and} \bibinfo{author}{\bibfnamefont{S.~D.~M.}
  \bibnamefont{{White}}}, \bibinfo{journal}{\prd}
  \textbf{\bibinfo{volume}{81}}, \bibinfo{eid}{083502}
  (\bibinfo{year}{2010}{\natexlab{a}}), \eprint{0910.5221}.

\bibitem[{\citenamefont{{Dent} et~al.}(2010)\citenamefont{{Dent}, {Dutta}, and
  {Scherrer}}}]{Dent_10}
\bibinfo{author}{\bibfnamefont{J.~B.} \bibnamefont{{Dent}}},
  \bibinfo{author}{\bibfnamefont{S.}~\bibnamefont{{Dutta}}}, \bibnamefont{and}
  \bibinfo{author}{\bibfnamefont{R.~J.} \bibnamefont{{Scherrer}}},
  \bibinfo{journal}{Physics Letters B} \textbf{\bibinfo{volume}{687}},
  \bibinfo{pages}{275} (\bibinfo{year}{2010}), \eprint{0909.4128}.

\bibitem[{\citenamefont{{Steigman} et~al.}(2012)\citenamefont{{Steigman},
  {Dasgupta}, and {Beacom}}}]{Steigman_12}
\bibinfo{author}{\bibfnamefont{G.}~\bibnamefont{{Steigman}}},
  \bibinfo{author}{\bibfnamefont{B.}~\bibnamefont{{Dasgupta}}},
  \bibnamefont{and} \bibinfo{author}{\bibfnamefont{J.~F.}
  \bibnamefont{{Beacom}}}, \bibinfo{journal}{\prd}
  \textbf{\bibinfo{volume}{86}}, \bibinfo{eid}{023506} (\bibinfo{year}{2012}),
  \eprint{1204.3622}.

\bibitem[{\citenamefont{{Watson}}(2010)}]{Watson_10}
\bibinfo{author}{\bibfnamefont{S.}~\bibnamefont{{Watson}}},
  \bibinfo{journal}{Perspectives On Supersymmetry II.~Series: Advanced Series
  on Directions in High Energy Physics, Edited by Gordon L Kane, vol.~21,
  pp.~305-324} \textbf{\bibinfo{volume}{21}}, \bibinfo{pages}{305}
  (\bibinfo{year}{2010}), \eprint{0912.3003}.

\bibitem[{\citenamefont{{Hooper}}(2013)}]{Hooper_13}
\bibinfo{author}{\bibfnamefont{D.}~\bibnamefont{{Hooper}}},
  \bibinfo{journal}{\prd} \textbf{\bibinfo{volume}{88}}, \bibinfo{eid}{083519}
  (\bibinfo{year}{2013}), \eprint{1307.0826}.

\bibitem[{\citenamefont{{Bringmann}}(2009)}]{Bringmann_09}
\bibinfo{author}{\bibfnamefont{T.}~\bibnamefont{{Bringmann}}},
  \bibinfo{journal}{New Journal of Physics} \textbf{\bibinfo{volume}{11}},
  \bibinfo{eid}{105027} (\bibinfo{year}{2009}), \eprint{0903.0189}.

\bibitem[{\citenamefont{{Kachelrie{\ss}} and {Serpico}}(2007)}]{Serpico_07}
\bibinfo{author}{\bibfnamefont{M.}~\bibnamefont{{Kachelrie{\ss}}}}
  \bibnamefont{and} \bibinfo{author}{\bibfnamefont{P.~D.}
  \bibnamefont{{Serpico}}}, \bibinfo{journal}{\prd}
  \textbf{\bibinfo{volume}{76}}, \bibinfo{eid}{063516} (\bibinfo{year}{2007}),
  \eprint{0707.0209}.

\bibitem[{\citenamefont{{Bergstr{\"o}m}
  et~al.}(2001)\citenamefont{{Bergstr{\"o}m}, {Edsj{\"o}}, and
  {Ullio}}}]{Bergstrom_01}
\bibinfo{author}{\bibfnamefont{L.}~\bibnamefont{{Bergstr{\"o}m}}},
  \bibinfo{author}{\bibfnamefont{J.}~\bibnamefont{{Edsj{\"o}}}},
  \bibnamefont{and} \bibinfo{author}{\bibfnamefont{P.}~\bibnamefont{{Ullio}}},
  \bibinfo{journal}{Physical Review Letters} \textbf{\bibinfo{volume}{87}},
  \bibinfo{eid}{251301} (\bibinfo{year}{2001}), \eprint{astro-ph/0105048}.

\bibitem[{\citenamefont{{Abdo} et~al.}(2010)\citenamefont{{Abdo}, {Ackermann},
  {Ajello}, {Atwood}, {Baldini}, {Ballet}, {Barbiellini}, {Bastieri},
  {Baughman}, {Bechtol} et~al.}}]{Abdo_2010}
\bibinfo{author}{\bibfnamefont{A.~A.} \bibnamefont{{Abdo}}},
  \bibinfo{author}{\bibfnamefont{M.}~\bibnamefont{{Ackermann}}},
  \bibinfo{author}{\bibfnamefont{M.}~\bibnamefont{{Ajello}}},
  \bibinfo{author}{\bibfnamefont{W.~B.} \bibnamefont{{Atwood}}},
  \bibinfo{author}{\bibfnamefont{L.}~\bibnamefont{{Baldini}}},
  \bibinfo{author}{\bibfnamefont{J.}~\bibnamefont{{Ballet}}},
  \bibinfo{author}{\bibfnamefont{G.}~\bibnamefont{{Barbiellini}}},
  \bibinfo{author}{\bibfnamefont{D.}~\bibnamefont{{Bastieri}}},
  \bibinfo{author}{\bibfnamefont{B.~M.} \bibnamefont{{Baughman}}},
  \bibinfo{author}{\bibfnamefont{K.}~\bibnamefont{{Bechtol}}},
  \bibnamefont{et~al.}, \bibinfo{journal}{Physical Review Letters}
  \textbf{\bibinfo{volume}{104}}, \bibinfo{eid}{101101} (\bibinfo{year}{2010}),
  \eprint{1002.3603}.

\bibitem[{\citenamefont{{Murase} and {Beacom}}(2012)}]{Murase_2012}
\bibinfo{author}{\bibfnamefont{K.}~\bibnamefont{{Murase}}} \bibnamefont{and}
  \bibinfo{author}{\bibfnamefont{J.~F.} \bibnamefont{{Beacom}}},
  \bibinfo{journal}{Journal of Cosmology and Astroparticle Physics}
  \textbf{\bibinfo{volume}{10}}, \bibinfo{eid}{043} (\bibinfo{year}{2012}),
  \eprint{1206.2595}.

\bibitem[{\citenamefont{{Zavala}
  et~al.}(2010{\natexlab{b}})\citenamefont{{Zavala}, {Springel}, and
  {Boylan-Kolchin}}}]{Zavala_10}
\bibinfo{author}{\bibfnamefont{J.}~\bibnamefont{{Zavala}}},
  \bibinfo{author}{\bibfnamefont{V.}~\bibnamefont{{Springel}}},
  \bibnamefont{and}
  \bibinfo{author}{\bibfnamefont{M.}~\bibnamefont{{Boylan-Kolchin}}},
  \bibinfo{journal}{\mnras} \textbf{\bibinfo{volume}{405}},
  \bibinfo{pages}{593} (\bibinfo{year}{2010}{\natexlab{b}}),
  \eprint{0908.2428}.

\bibitem[{\citenamefont{{Slatyer} et~al.}(2009)\citenamefont{{Slatyer},
  {Padmanabhan}, and {Finkbeiner}}}]{Slatyer_09}
\bibinfo{author}{\bibfnamefont{T.~R.} \bibnamefont{{Slatyer}}},
  \bibinfo{author}{\bibfnamefont{N.}~\bibnamefont{{Padmanabhan}}},
  \bibnamefont{and} \bibinfo{author}{\bibfnamefont{D.~P.}
  \bibnamefont{{Finkbeiner}}}, \bibinfo{journal}{\prd}
  \textbf{\bibinfo{volume}{80}}, \bibinfo{eid}{043526} (\bibinfo{year}{2009}),
  \eprint{0906.1197}.

\bibitem[{\citenamefont{{Madhavacheril}
  et~al.}(2013)\citenamefont{{Madhavacheril}, {Sehgal}, and
  {Slatyer}}}]{Slatyer_13}
\bibinfo{author}{\bibfnamefont{M.~S.} \bibnamefont{{Madhavacheril}}},
  \bibinfo{author}{\bibfnamefont{N.}~\bibnamefont{{Sehgal}}}, \bibnamefont{and}
  \bibinfo{author}{\bibfnamefont{T.~R.} \bibnamefont{{Slatyer}}},
  \bibinfo{journal}{ArXiv e-prints}  (\bibinfo{year}{2013}),
  \eprint{1310.3815}.

\bibitem[{\citenamefont{{Ando}}(2009)}]{Ando_09}
\bibinfo{author}{\bibfnamefont{S.}~\bibnamefont{{Ando}}},
  \bibinfo{journal}{\prd} \textbf{\bibinfo{volume}{80}}, \bibinfo{eid}{023520}
  (\bibinfo{year}{2009}), \eprint{0903.4685}.

\bibitem[{\citenamefont{{Sanchez-Conde} and {Prada}}(2013)}]{MASC_13}
\bibinfo{author}{\bibfnamefont{M.~A.} \bibnamefont{{Sanchez-Conde}}}
  \bibnamefont{and} \bibinfo{author}{\bibfnamefont{F.}~\bibnamefont{{Prada}}},
  \bibinfo{journal}{ArXiv e-prints}  (\bibinfo{year}{2013}),
  \eprint{1312.1729}.

\end{thebibliography}



\end{document}